\documentclass[%
 reprint,
nofootinbib,
 amsmath,amssymb,
 aps,
superscriptaddress
]{revtex4-2}
\usepackage{graphicx}
\usepackage{dcolumn}
\usepackage{bm}
\usepackage[hidelinks]{hyperref}
\usepackage{color}
\usepackage{soulutf8}
\usepackage{siunitx}
\usepackage{cleveref}
\usepackage{comment}
\usepackage[normalem]{ulem}

\begin{document}

\preprint{APS/123-QED}

\title{Axion dark matter detection via shift current} 

\author{Dan Kondo}%
 \email{dan.kondo@ipmu.jp}
 \affiliation{Kavli Institute for the Physics and Mathematics of the Universe (WPI), University of Tokyo Institutes for Advanced Study, University of Tokyo, Kashiwa 277-8583, Japan
}%
\author{Takahiro Morimoto}
 \email{morimoto@ap.t.u-tokyo.ac.jp}
 \affiliation{Department of Applied Physics, The University of Tokyo, Tokyo 113-8656, Japan}
\author{Genta Osaki}
\email{osaki@hep-th.phys.s.u-tokyo.ac.jp}
\author{Thanaporn Sichanugrist}
\email{thanaporn@hep-th.phys.s.u-tokyo.ac.jp}
\affiliation{%
 Department of Physics, The University of Tokyo, Tokyo 113-8656, Japan
}%

\date{\today}

\begin{abstract}

We propose a novel method to detect axion dark matter based on a topological phenomenon known as the shift current. We make use of the second-order nonlinearity of the shift current by applying a strong oscillating electric field. This field enhances the axion-induced shift current signal and downconverts its frequency to a more accessible range. The nondissipative nature of the shift current allows us to achieve broadband detection via difference frequency generation.  We demonstrate the possibility of probing QCD axions and axion-like particles using the properties of the type-I Weyl semimetal TaAs, targeting an axion mass of $O(10)\,\mathrm{meV}$, corresponding to a photon coupling of $g_{a\gamma\gamma} \simeq \mathcal{O}(10^{-11})\,\mathrm{GeV}^{-1}$.

\end{abstract}

\maketitle


\textit{Introduction ---}
The identification of dark matter is a long-standing problem in particle physics, cosmology, and astrophysics. Its particle-physics properties, such as its mass and the coupling to the standard model particles, are largely unknown. One of the promising candidates for dark matter is the axion or axion-like particle (ALP). An axion is a Nambu-Goldstone boson arising from the global $\mathrm{U}(1)$ Peccei-Quinn symmetry \cite{Weinberg:1977ma,Wilczek:1977pj} that solves the strong CP problem \cite{Peccei:1977hh}. 
It is expected that axions or ALPs could have a feeble coupling to photons \cite{Witten:1984dg,Conlon:2006tq,Svrcek:2006yi,Arvanitaki:2009fg,Cicoli:2012sz}. (For comprehensive reviews, see \cite{Kim:2008hd,Arias:2012az,Kawasaki:2013ae,Marsh:2015xka,DiLuzio:2020wdo,OHare:2024nmr,Ferreira:2020fam,Safdi:2022xkm}). At present, one of the most promising approaches to look for the axion dark matter is to utilize the axion-photon conversion in external electric or magnetic fields \cite{Sikivie:1983ip}. In particular, cavity haloscope experiments utilize the cavity resonance applied with an external strong magnetic field to detect the axion-induced photons~\cite{PhysRevLett.51.1415}. This has great sensitivity mainly at the dark matter mass range $\mathcal{O}(1)$ - $\mathcal{O}(100) \ \rm \mu eV$. 
However, at higher axion masses, the corresponding signal frequencies exceed the typical resonant frequency of the conventional cavity, and such a region remains to be probed.

In this article, we propose a novel method for broadband detection of axion dark matter of relatively high mass $\mathcal{O}(100)$ meV using a topological phenomenon called ``shift current''.
The shift current is a nonlinear optical response of inversion-broken materials where a light-induced direct current appears without the need for conventional semiconductor junctions~\cite{PhysRev.127.2036,fridkin1967current,Koch01011976,ghosez1995born,PhysRevB.96.241203}. 
The shift current arises from a spatial shift in the quantum wave function of electronic states during state transitions.
Owing to its quantum nature, the output shift current exhibits high tolerance to defects, while offering a fast temporal response~\cite{doi:10.1073/pnas.2007002117}.
It has been applied in studies of terahertz (THz) dynamics~\cite{sotome2018spectral}, corresponding to physics with the energy scale of meV and higher.

Our axion dark matter search idea with the shift current is as follows.
First, we apply a static magnetic field to convert axions to an electric field. Second, we externally apply another strong oscillating electric field, which hereafter we also call the experimental field. Since the shift current is a nonlinear response, the two electric fields together can interact with material and induce an enhanced axion shift current to be observed. The frequency of the resulting current is determined by the frequency difference of the two electric fields. Therefore, when the high frequency of the axion-induced electric field aligns with or around that of the experimental field, a DC or lower-frequency signal is generated, which is more readily detectable. In this way, by changing the frequency of the experimental field to several values, one may search for the axion dark matter signal in a desired mass range.   Importantly, the non-dissipative nature of the shift current response enables broadband detection through difference frequency generation, maintaining stable conductivity without being suppressed by electron scattering effects. 
In particular, with one fixed frequency of the experimental field, the simultaneously detectable frequency of the output current is determined by the band structure of the material and the frequency of the experimental field itself, which generally can be broader than the conventional resonant experiments.
Based on our method, we demonstrate that, using the type-I Weyl semimetal TaAs, it is possible to probe the parameter space of QCD axions and ALPs, targeting a mass of $O(10)\,\mathrm{meV}$ 
with the output signal frequency reduced to DC or below $\mathcal{O}(1)\,\mathrm{THz}$.

\textit{Axion-induced shift current ---}\label{sec:AxionDetection}
In the presence of axion dark matter under a magnetic field, the axion can induce the electric field through axion-photon coupling given by
\begin{align}
    \mathcal{L}\supset -\frac{g_{a\gamma\gamma}}{4}aF_{\mu\nu}\tilde{F}^{\mu\nu}
    =g_{a\gamma\gamma}a\bm{E}\cdot\bm{B}, \label{eq:gagamma}
\end{align}
where $a$ is the axion field and $F$ is the electromagnetic field strength, $\tilde{F}^{\mu\nu}=\epsilon^{\mu\nu\rho\sigma}F_{\rho\sigma}/2$ with the totally antisymmetric tensor $\epsilon^{\mu\nu\rho\sigma}$ (with $\epsilon^{0123}=1$). For the QCD axion model, the coupling constant is given by \cite{GrillidiCortona:2015jxo}
\begin{align}
    g_{a\gamma\gamma}
    &=
    \frac{\alpha}{2\pi f_a}\left(\frac{E}{N}-1.92\right)
    =
    \left(0.203\frac{E}{N}-0.39\right)\frac{m_a}{\text{GeV}^2},
\end{align}
where $m_a$ is the axion mass. $E$ and $N$ are the electromagnetic and color anomalies associated with axion axial current, which takes a variety of values \cite{Zhitnitsky:1980tq,Dine:1981rt,Kim:1979if,Shifman:1979if,Kim:1998va,DiLuzio:2016sbl,DiLuzio:2017pfr}, and $E/N$ are from $5/3$ to $44/3$. Other than those, there are models whose values are different from the band \cite{Farina:2016tgd,Agrawal:2017cmd,Hook:2018jle,DiLuzio:2021pxd,Sokolov:2021ydn,Diehl:2023uui}. For ALP, however, $g_{a\gamma\gamma}$ becomes a free parameter.
We assume dark matter to consist only of axions and focus on axions of mass $\sim\mathcal{O}(10)$ - $\mathcal{O}(100) \ \rm meV$. To satisfy the observed energy density of dark matter $\rho_{\rm DM}$, the number of dark matter particles within the de Broglie wavelength is enormous; their wave functions overlap and can be treated as a classical wave \cite{Foster:2017hbq}. Then, the amplitude of the axion field is given by
\begin{equation}
    a(t)=\frac{\sqrt{2 \rho_{\rm DM}}}{m_\mathrm{DM}}\cos(m_\mathrm{DM} t+\phi_{\rm DM}),
\end{equation}
where $m_{\mathrm{DM}}$ is the dark matter mass ($=m_a$) and $\phi_{\mathrm{DM}}$ is the phase of the dark matter field.
When we apply a static magnetic field $B_0$, the electric field is induced by the axion through the interaction given by Eq.~\eqref{eq:gagamma}; it has the same direction as the magnetic field, while the amplitude is given by 
\begin{align}\label{eq:Eaxion}
E_{\text{DM}}
&=
g_{a\gamma\gamma}B_0\frac{\sqrt{2\rho_{\text{DM}}}}{m_\mathrm{DM}}\nonumber\\
&\simeq
7.9\times 10^{-12} \frac{\text{V}}{\text{m}} \left( \frac{g_{a\gamma\gamma}}{10^{-11} \ \text{GeV}^{-1}} \right) \left(\frac{B_0}{1 \ \text{T}} \right) \nonumber\\
&\;\;\;\;\;\times
\left(\frac{m_\mathrm{DM}}{1\ \text{meV}}\right)^{-1}\left(\frac{\rho_{\text{DM}}}
{0.45\ \text{GeV}\text{cm}^{-3}}\right)^{\frac{1}{2}}.  \end{align}

The shift current is regarded as a second-order DC response and can be expressed in terms of the integral over the shift of the Berry connection over the Brillouin zone and is directly related to the shift of the center of mass of electron states \cite{Morimoto_2023}. Indeed, the same effect can also be observed with a small frequency current as well \cite{de_Juan_2020}. In this article, we also call it simply shift current and will apply it for axion signal broadband search.
By formulating a coherent state path integral formalism for an electronic system at finite temperature, we derive the general response up to the second order with general electric fields. Details of the derivation are provided in Appendix~\ref{sec:PathIntegral}-\ref{sec:Keldysh}. With general input electric field $\bm{E}(t)=\int d\omega \ e^{-i \omega t} \bm{E}(\omega)$, the shift current with small frequency $\omega$ is given by
\begin{align}
\langle&\widehat{J}_{\rm shift}^{\mu}\rangle^{(2)}(\omega)
    =\;
    \frac{1}{2}
    \sum_{ab}
    \int
    \frac{d^{d}\bm{k}}{(2\pi)^{d}}
    \int d\omega_{1}d\omega_{2}
    \left(
    \frac{-e^{3}}{\omega_{1}\omega_{2}}
    \right)
    \nonumber\\
    &\times
    E^{\alpha_{1}}(\omega_{1})E^{\alpha_{2}}(\omega_{2})
    \delta(\omega-\omega_{1}-\omega_{2})
    \nonumber\\
& \times r^{\mu\alpha_{1}}_{\bm{k}ab}r^{\alpha_{2}}_{\bm{k}ba}
 \left(\frac{f_{ab } }{\omega_2-\varepsilon_{\bm{k}ba} +i\gamma} \right) + (\alpha_1,\omega_1 \leftrightarrow \alpha_2,\omega_2),\label{eq:shift}
\end{align}
where
\begin{gather}
r^{\mu}_{\bm{k}ab}=i \varepsilon_{\bm{k}ab} \mathcal{A}_{ab}^{\mu}(\bm{k}),\\
r^{\mu\alpha}_{\bm{k}ab}=\left[\partial_{k_\mu} r^\alpha_{\bm{k}ab} -i \left(\mathcal{A}^\mu_{aa}(\bm{k})-\mathcal{A}^\mu_{bb}(\bm{k})\right)r^\alpha_{\bm{k}ab}\right],
\end{gather}
with the Berry connection defined by
\begin{equation}
    \mathcal{A}^\mu_{ab}(\bm{k})\equiv i\langle u_{\bm{k}a}|\frac{\partial}{\partial{k_\mu}}|u_{\bm{k}b}\rangle_{V}. \label{eq:berrydef}
\end{equation}
The integration over $\bm{k}$ in Eq.~\eqref{eq:shift} is for the Brillouin zone.
The state $|u_{\bm{k}a}\rangle$ is the Bloch state of the energy level $a$ with momentum $\bm{k}$, $\varepsilon_{\bm{k}ab}\equiv\varepsilon_{\bm{k}a}-\varepsilon_{\bm{k}b}$ with $\varepsilon_{\bm{k}a}$ denoting the energy level $a$ with momentum $\bm{k}$, and $f_{ab}\equiv f_a-f_b$ with $f_a=f(\varepsilon_{\bm{k}a})$ denoting the Fermi-Dirac distribution of electron in $\varepsilon_{\bm{k}a}$. The subscript $V$ in Eq.~\eqref{eq:berrydef} means that the spatial integration over the wave function is performed within one unit cell. The energy conservation law is reflected in the delta function $\delta(\omega-\omega_{1}-\omega_{2})$. Phenomenologically, to take into account the electron scattering effect in the material, we replaced
$
\omega_{i}$ for $\omega_{i}+ i\gamma,
$
where $\gamma$ is determined by the scattering rate \cite{parker2019diagrammatic}.
To derive Eq.~\eqref{eq:shift}, we also assumed $\gamma,\omega$ are sufficiently small compared to the energy scale of the input field.
Throughout this article, we consider a time-reversal symmetry material with a broken spatial-inversion symmetry. 
The nonzero shift current can be obtained provided that there is $\bm{k}$ in the Brillouin zone satisfying $\varepsilon_{\bm{k}ba}\simeq \omega_{1,2}$ for an arbitrary pair $(a,b)$. Then, the resonant transition occurs, generating a shift current. 
With the time-reversal symmetry condition, the intrinsic second-order response other than the shift current is purely circular and becomes zero for a linearly polarized field. We assume that the shift current (Eq.~\eqref{eq:shift}) is the dominant contribution among the second-order processes. 

In our strategy, to enhance and detect a feeble dark matter field, we apply another electric field $E_{\mathrm{exp}}$ and detect the cross-term of the shift-current response ($J\propto E_{\mathrm{DM}}E_{\mathrm{exp}}$). Then, a sizable output current is generated by applying high power $E_{\mathrm{exp}}$.
In total, the external electric field is given by
\begin{align}
    E(t)
    =\mathrm{Re}\left[
    E_{\mathrm{DM}}
    e^{im_{\mathrm{DM}}t-i\phi_{\mathrm{DM}}}
    +
    E_{\mathrm{exp}}
    e^{-i\omega_{\mathrm{exp}}t}\right],
    \label{eq:external_E}
\end{align}
where, for convenience, we set the sign of the exponent oppositely and assume $m_{\rm DM},\omega_{\rm exp}>0$ without loss of generality. Then, the relevant signal shift current has frequency $\Delta\omega=-m_{\rm DM}+\omega_{\rm exp}$.
We read out the current of the form :
\begin{align}
    \langle\widehat{J}^{\mu}\rangle_{\mathrm{shift}}^{(2)}(\Delta\omega)=\sigma^{\mu\alpha_{1}\alpha_{2}}_{\mathrm{shift}}(\Delta\omega)E^{\alpha_{1}}_{\mathrm{DM}}e^{-i\phi_{\mathrm{DM}}}
    E^{\alpha_{2}}_{\mathrm{exp}},
\end{align}
where $\sigma^{\mu\alpha_{1}\alpha_{2}}_{\mathrm{shift}}(\Delta\omega)$ is the shift current conductivity.
The conductivity reads 
\begin{align}
&\sigma_{\rm shift}^{\mu\alpha_1 \alpha_2}(\Delta \omega) = \sum_{ab}
    \int
    \frac{d^{d}\bm{k}}{(2\pi)^{d}}
    \left(
    \frac{e^{3}}{m_{\rm DM}\omega_{\rm exp}}
    \right)
\nonumber\\
&\times f_{ab}\left[\frac{r^{\mu\alpha_1}_{\bm{k}ab}r^{\alpha_2}_{\bm{k}ba}}{m_{\rm DM}-\varepsilon_{\bm{k}ba}+i\gamma}-\frac{r^{\mu\alpha_{2}}_{\bm{k}ba}r^{\alpha_1}_{\bm{k}ab}}{\omega_{\rm exp}- \varepsilon_{\bm{k}ba}+i\gamma}\right]. \label{eq:sigmadmexp}
\end{align}
The typical energy scale of electronic bands is $\sim \mathcal{O}(100) \ \rm THz$. We consider applying $\omega_{\rm exp} \sim \mathcal{O}(10) \ \rm THz$ and focus on current output with frequency $\Delta \omega \lesssim \mathcal{O}(100) \ \rm GHz$. With this condition, the resonant transition of electrons can occur, and Eq.~\eqref{eq:sigmadmexp} is well approximated. Also, the signal can be easily detected compared to the original source.
The signal linewidth is determined by that of $E_{\mathrm{DM}}$ and $E_{\mathrm{exp}}$.
We take the linewidth of $E_{\rm exp}$ and $E_{\rm DM}$ as $\Delta \omega_{\rm exp} =10 \ \rm GHz$~\cite{lin2022over} and $\Delta m_{\rm DM}\simeq m_\mathrm{DM} v_{\rm DM}^2 \simeq  0.1 \ \rm GHz $, respectively, where we assume dark matter velocity $v_{\rm DM} \simeq 10^{-3}$.
Therefore, we expect the linewidth of the signal to be $\sigma_{\rm sig} \sim \max\{\Delta m_{\rm DM},\Delta \omega_{\rm exp} \}\simeq 10 \ \rm GHz$ due to the convolution of two input fields. 
\begin{figure}[tbp]
    \centering
    \includegraphics[width=0.95
    \linewidth]{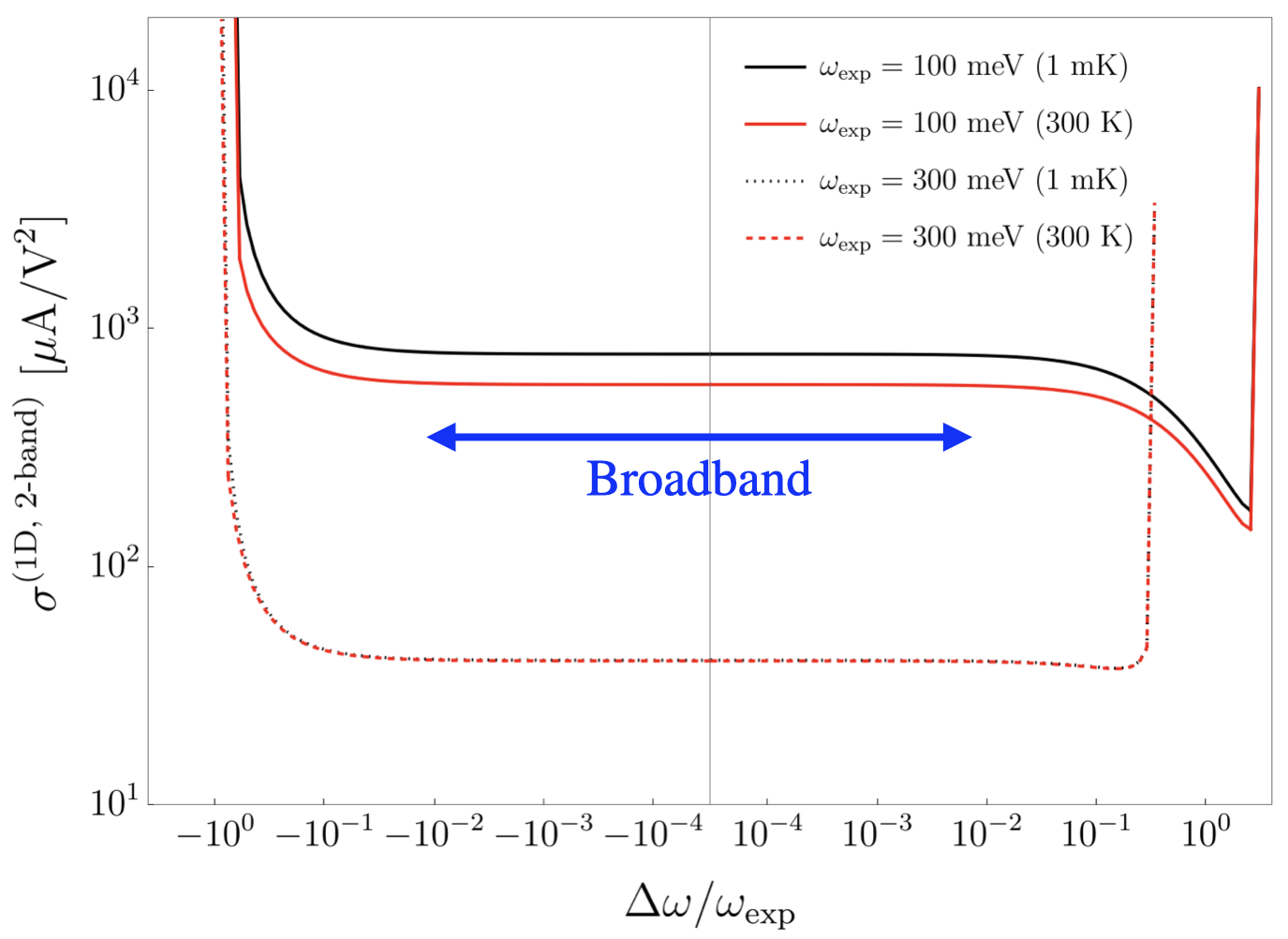}
    \caption{Plot of the shift-current conductivity as a function of $\Delta \omega/\omega_{\rm exp}$ with fixed frequency $\omega_{\text{exp}}$ of the 1D Rice-Mele model. We illustrate the characteristics of the response with two fields of the 2-band system. We set the energy gap covering 30 meV to 400 meV depending on momentum $k$. 
    The conductivities are derived with $T=1 \ \rm mK$ and $T=300 \ \rm mK$ with the Fermi energy lying at the middle of the two bands. 
    The divergent values on the edge come from the points where the group velocity vanishes. }
    \label{fig:Domega}
\end{figure}

Our method for axion search can be performed in a broadband manner.
We illustrate this property considering conductivity $\sigma^{(\rm{1D,2band})}_{\mathrm{shift}}$ in a two-band system with a 1D Rice-Mele Hamiltonian \cite{RiceMele1982} as a toy model (see Appendix~\Ref{sec:Ricemale} for details). The system consists of two types of ions as a unit cell aligned repeatedly in a 1-dimensional line, and the electron state is described by a tight-binding model. The model has two energy bands of electron states, where we set an energy gap covering 30-400 meV depending on momentum $k$, while the spatial inversion symmetry is broken by a difference in the electron potential between two ions in the unit cell.  
The shift-current conductivity in this model as a function of $\Delta \omega/\omega_{\rm exp}$ with fixed frequency $\omega_{\text{exp}}$ is shown in Fig.~\ref{fig:Domega}. In the plot, we directly use Eq.~\eqref{eq:sigmadmexp}, however, it should be kept in mind that the equation is reliable for $\Delta\omega \ll \omega_{\rm exp}$.
Note that the response amplitude remains nearly constant for small frequency deviations $\Delta \omega \ll  \omega_{\rm exp}$.
Also, the plausible width of $\Delta \omega$ for axion search is not determined by the scattering rate $\gamma$ as in the usual resonance experiment, but directly by the band structure. This is because the resonant transition is always automatically ensured by integration over $\bm{k}$ in Eq.~\eqref{eq:sigmadmexp} provided that the source has a frequency that matches the band gap for an arbitrary $\bm{k}$ in the Brillouin zone.

\begin{figure}[tbp]
    \centering
    \includegraphics[width=1\linewidth]{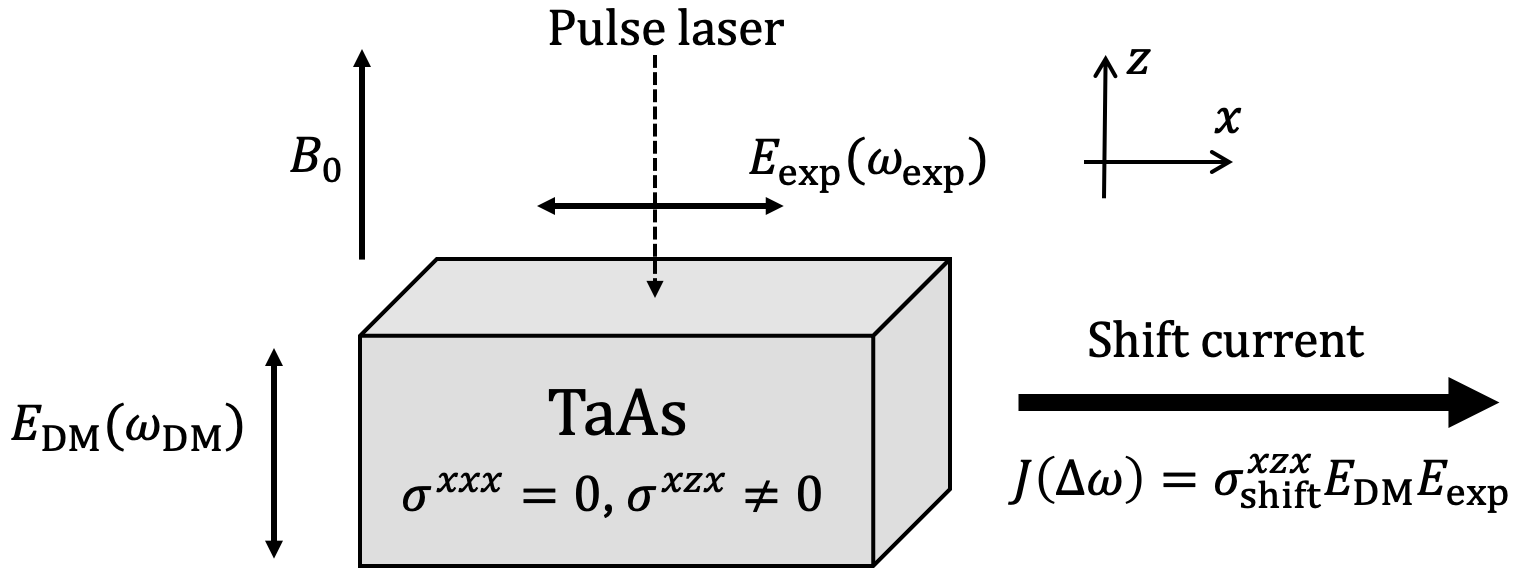}
    \caption{Schematic picture of the experimental setup. The static magnetic field $B_{0}$ converts an axion to a photon $E_{\mathrm{DM}}$. In a Type-I Weyl semimetal TaAs, the second order response of $E_{\mathrm{DM}}$ and $E_{\mathrm{exp}}$ generates the output signal $J$. We look for the axion dark matter satisfying the condition $\Delta\omega=-m_{\mathrm{DM}}+\omega_{\mathrm{exp}}$ where $\Delta\omega$ is a small frequency difference. Since $\sigma^{xxx}=0,\;\sigma^{xzx}\neq0$, the background  $J_{\mathrm{BG}}=\sigma^{xxx}E_{\mathrm{exp}}E_{\mathrm{exp}}$ vanishes.}
    \label{fig:setup}
\end{figure}

We exemplify TaAs as a real material setup. TaAs is a Type-I Weyl semimetal that lacks inversion symmetry and supports a bulk photovoltaic effect.
In a 3D system, the breaking of spatial inversion symmetry can generally lead to the appearance of degenerate Weyl nodes that host gapless excitations with topological stability \cite{nagaosa2020transport}. 
Since TaAs has $C_{4v}$ symmetry, the $xxx$ component of the current conductivity $\sigma^{xxx}$ is identically zero due to the mirror symmetry with the $yz$ plane, while the $xzx$ component $\sigma^{xzx}$ has a nonzero contribution \cite{sturman_photovoltaic_1992}.
This directional property of the material provides the great benefit that we can carefully choose the direction of the input electric field such that the background current from the experimental field $J_{\mathrm{BG}} \propto E_{\mathrm{exp}}E_{\mathrm{exp}}$ vanishes. 
In Fig.~\ref{fig:setup}, we show the schematic picture of the experimental setup. We apply a static magnetic field $B_{0}$ in the $z$ direction, which converts the axion to a photon $E_{\mathrm{DM}} \parallel z$. We also apply an oscillating electric field $E_{\mathrm{exp}}$ in the $x$ direction. Measuring the current in the $x$ direction, we detect the axion-induced shift current $J=\sigma^{xzx}_{\mathrm{shift}} E_{\mathrm{DM}}E_{\mathrm{exp}}$. In this setup, the background second-order response $J_{\mathrm{BG}}=\sigma^{xxx} E_{\mathrm{exp}}E_{\mathrm{exp}}$ vanishes.  One should also consider the general background other than the second-order response that might be generated from the experimental field $E_{\rm exp}$. We note that the mirror symmetry with the $yz$ plane also forbids all of the even-order responses in the $x$ direction, so we can neglect the background with a frequency similar to that of the signal. On the other hand, the other currents from general odd-order responses, have a frequency of at least $\omega_{\rm exp}$, which is far from the signal, so they can also be neglected.

Later in the subsequent sensitivity analysis for TaAs, we consider small $\Delta \omega$ within $\Delta\omega \lesssim \omega_{\rm exp}/100$ and estimate shift current conductivity using DC shift current value.
The shift-current conductivity for TaAs is observed to be $\sigma^{xzx}_{\mathrm{shift}}\sim 200~\mathrm{\mu A/V^2}$ with an input field with frequency around 100 meV, while it is expected to be in the range $100-2500~\mathrm{\mu A/V^2}$ with the input field's frequency range $25$-$350 \ \rm meV$ (see \cite{osterhoudt2019colossal} and its Supplemental Material). The peak value $O(1000)~\mathrm{\mu A/V^2}$ corresponds to the energy around $25$ - $50 \ \rm meV$. For later analysis, we set the typical value $\sigma^{xzx}_{\mathrm{shift}}\sim 200~\mathrm{\mu A/V^2},~2000~\mathrm{\mu A/V^2}$ according to the scanning mass range. 

\textit{Protocol \& Sensitivity ---}\label{subsec:sensitivity}
We need to take into account the effect of the dark matter phase $\phi_{\mathrm{DM}}$.
We set the measurement time $\tau$ to be $\min\{1/\Delta \omega_{\rm exp},1/\Delta m_{\rm DM} \} \simeq 0.1 \ \rm ns$ so that, within one cycle of measurement, the signal coherently oscillates and the phase $\phi_{\mathrm{DM}}$ can be treated as constant (but random for each measurement). 
In addition, dark matter also has spatial fluctuations and the corresponding coherence length scale is $\lambda_{\rm De-Broglie}= 1/(m_{\rm DM}v_{\rm DM})\simeq \mathcal{O}(1) \ \rm cm$. We then set an experimental length scale $L_{\rm exp}$ to be $\mathcal{O}(1) \ \rm cm$.

The following is the procedure for signal collection for sensitivity estimation:
\begin{enumerate}
\item We apply an oscillating electric field $E_{\rm exp}$ and magnetic field $B_0$ to the system and record the time evolution of the output current. The signal current is to be picked up using the conductor pads connected to the
two ends of the sample, transmitted through a coaxial cable, and analyzed with
a spectrum analyzer \cite{keysight8566b}.
\item For each time interval $\tau=0.1~\mathrm{ns}$, we Fourier transform the data and read out signal power $P_{\rm sig}$ at the chosen frequency $\Delta \omega$. 
We take the resolution bandwidth or bin's width as $\Delta f_{\rm bin}\sim \sigma_{\rm sig}= 10\ \rm GHz$. 
Importantly, we can probe many bins simultaneously; the detectable response frequency $\Delta \omega$ is broadband covering from 0 Hz to  100 GHz. We denote this range as $\Delta f_{\rm scan}=100 \ \rm GHz$ and the interrogation time for this range as $\tau_{\rm scan}$.

\item We change $\omega_{\rm exp}$ to several values or sweep it during the experiment so that the desired scan range of $\omega_{\rm DM}$ is achieved. Using several values of $\omega_{\rm exp}$ is necessary to distinguish the sign of $\Delta\omega$.
\end{enumerate}
The theoretical prediction of the averaged current signal power for each bin is
\begin{align}
    \langle
    &P_{\mathrm{sig}}
    \rangle
    =
    \frac{1}{2}
    R_{\mathrm{exp}}|\sigma^{xzx}_{\mathrm{shift}}|^{2}E_{\mathrm{DM}}^{2}E_{\mathrm{exp}}^{2}L_{\mathrm{exp}}^{4}\nonumber\\
    &=6.2\times 10^{-15}~\mu\mathrm{W}
    \left(\frac{R_{\mathrm{exp}}}{50~\Omega}\right)
    \left(\frac{\sigma^{xzx}_{\mathrm{shift}}}{200~\mu\mathrm{A/V}^{2}}\right)^{2}\nonumber\\
    &\;\;\;\;\;\times
    \left( \frac{g_{a\gamma\gamma}}{10^{-11} \ \text{GeV}^{-1}} \right)^{2}
    \left(\frac{B_0}{1 \ \text{T}} \right)^{2}
    \left(\frac{m_\mathrm{DM}}{1\ \text{meV}}\right)^{-2}\nonumber\\
    &\;\;\;\;\;\times
    \left(\frac{\rho_{\text{DM}}}{0.45\ \text{GeV}\text{cm}^{-3}}\right)
    \left(\frac{E_{\text{exp}}}{10^8\ \text{V}/ \text{m}}\right)^{2}
    \left(\frac{L_{\text{exp}}}{1\ \text{cm}}\right)^{4},
\end{align}
where $R_{\mathrm{exp}}$ is the resistance of the conductor for collecting the current and $L_{\mathrm{exp}}$ is the length scale of the sample. Note that the randomness of the phase $\phi_{\rm DM}$ of each measurement is taken into account as the factor 1/2 in front.
We suppose that the main noise comes from Johnson-Nyquist noise \cite{johnson1928thermal,nyquist1928thermal}. Then, the signal-to-noise ratio (SNR) for each bin of $m_{\rm DM}$ is obtained by the radiometer equation \cite{10.1063/1.1770483} as
\begin{align}
    \mathrm{SNR}
    =
    \frac{\langle P_{\mathrm{sig}}\rangle}{4k_{\mathrm{B}}T }
    \sqrt{\frac{\tau_{\rm scan} }{\Delta f_{\mathrm{bin}}}},
    \label{eq:SNR}
\end{align}
where 
$\tau_{\rm scan}\equiv T_{\rm total}\times ({\Delta f_{\rm scan}}/{\Delta f_{\rm total}})\times D$ is the time for a single scan with $T_{\rm total}$ being the total observation time, $\Delta f_{\rm total}$ the total scan range, and $D$ the duty cycle chosen such that the heat load (assumed equal to an applied power) does not exceed the cooling capacity.
Here, we assumed $4.2~\rm K$ thermal setup, with a cooling rate of $5~{\rm W}$, which is within the performance range of the existing two-stage pulse tube cryocooler~\cite{Hao_2024}.

\begin{figure}[tbp]
    \centering
    \includegraphics[width=0.95\linewidth]{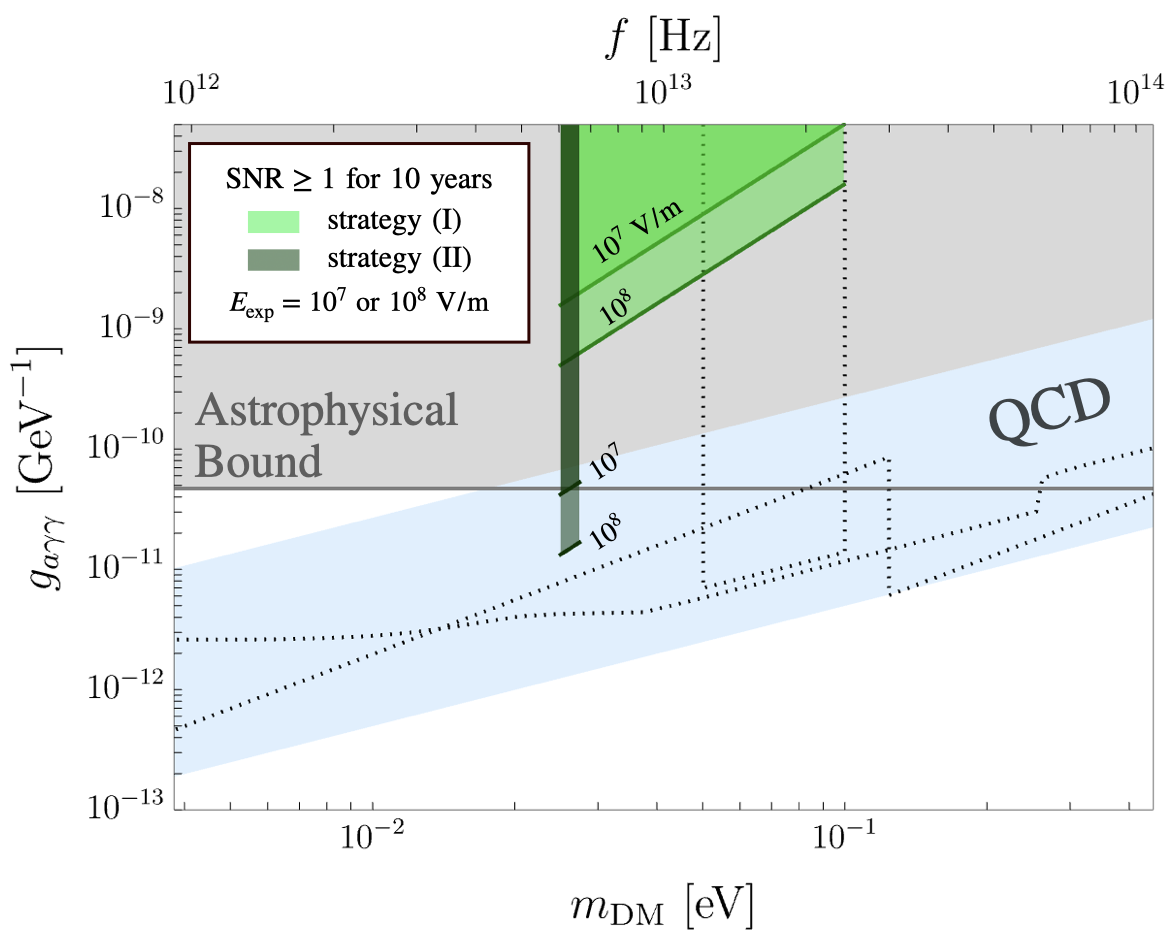}
    \caption{Sensitivity to the axion photon coupling $g_{a\gamma\gamma}$ determined by the SNR assuming TaAs as a sample. The dotted lines represent the other experimental proposals (IAXO \cite{Irastorza:2011gs,IAXO:2019mpb,Ge:2020zww}, BREAD \cite{BREAD:2021tpx} and WISPFI 
 \cite{Batllori:2023gwy}). The blue band corresponds to the allowed region of the QCD axion. We assumed $B_0=10 \ \rm T$ and $E_{\rm exp}=10^{7},~10^{8} \ \rm V/m$ with $5~{\rm W}/4.2~\rm K$ two-stage pulse tube cryocooler~\cite{Hao_2024}. The total observation time is 10 years.}
    \label{fig:Axionplot}
\end{figure}
We show the expected sensitivity of TaAs determined by $\mathrm{SNR}\geq1$ in Fig.~\ref{fig:Axionplot} where we used $B_0=10 \ \rm T$, $E_{\mathrm{exp}}=10^7,~10^{8}~\mathrm{V/m}$ (referenced from the experiment applying THz pulse laser \cite{sanari2020modifying}, where the former value represent a more conservative choice), $T=4.2~\mathrm{K}$, $R_{\mathrm{exp}}=50~\Omega$, $\Delta f_{\mathrm{bin}}=10~\mathrm{GHz}$, and $\Delta f_{\mathrm{scan}}=100~\mathrm{GHz}$.
Strategy (I) corresponds to the wide scanning in the range of 25-100 meV, while strategy (II) concentrates on a single scan range around $25\ {\rm meV}$. We used $\sigma^{xzx}_{\mathrm{shift}}= 200~\mathrm{\mu A/V^2}$ for strategy (I) and $ 2000~\mathrm{\mu A/V^2}$ for strategy (II) \cite{osterhoudt2019colossal}. The total observation time is 10 years. To solve the equation of ${\rm SNR=1}$, we took into account the maximum sample size scale limited by the coherence length of axion DM.
The gray dotted lines represent the prospects of IAXO \cite{Irastorza:2011gs,IAXO:2019mpb,Ge:2020zww}, BREAD \cite{BREAD:2021tpx} and WISPFI \cite{Batllori:2023gwy} (for other experiments, see \cite{Baryakhtar2018,Chiles2022,Manenti2022,Janish2023}). The blue band corresponds to the allowed region of the QCD axion. As a result, QCD axions may be probed by strategy (II) when focusing at frequency with sufficiently large conductivity ($\sigma= 2000 \ \rm \mu A/ V^2$). In strategy (I), the lower conductivity is insufficient to surpass current astrophysical limits. 

\textit{Conclusion and discussion ---}\label{sec:Conclusion}
In this work, we have proposed a novel method to probe axion dark matter using the shift current in topological materials. We demonstrate that, with an input oscillating electric field of $10^7\sim10^8\,\mathrm{V/m}$ and a static magnetic field of $10\,\mathrm{T}$, QCD axions and ALPs might be probed using the type-I Weyl semimetal TaAs, targeting a mass of $25\,\mathrm{meV}$ within a relative mass window of $\pm 1\%$
($\sim100 \ \mathrm{GHz}$), 
down to a photon coupling of $g_{a\gamma\gamma} \simeq \mathcal{O}(10^{-11})\,\mathrm{GeV}^{-1}$, with a total observation time of 10 years. 
We note that comparable sensitivity can also be achieved when probing axion masses in the range of $20$--$30 \ \mathrm{meV}$ owing to the similar conductivity~\cite{osterhoudt2019colossal}. On the other hand, the scanning of the entire material bandwidth does not yield sufficient sensitivity where the conductivity is too small. 

In this study, we assumed $R_{\rm exp}=50 \ \rm \Omega$ load for the sensitivity estimation based on the typical termination of the readout system, but the internal resistance of the sample could be made larger for disordered samples (leading to large open circuit voltage due to shift current which tends not to be suppressed by disorder~\cite{doi:10.1073/pnas.2007002117}), thereby possibly further boosting the exported signal power relative to our baseline.

We showed that, in principle, applying electric and magnetic fields along the direction dictated by the mirror symmetry of the material can eliminate the background current generated from the application of the oscillating electric field. We have neglected the fluctuations of the experimental field in the unintended direction; indeed, such fluctuations could still cause the background current. 
Since this background contribution is essentially a DC component, it may be distinguished from signals at nonzero frequencies via difference-frequency generation~\cite{de_Juan_2020}. A quantitative verification of this effect remains as a subject for future work.

A practical challenge lies in the application of the strong experimental electric field. The assumed oscillating field strength for the signal enhancement, on the order of $10^{8} \ \rm V/m$, is currently achievable using short-pulse laser technology with its beam size $O(10)~{\rm \mu m}$ \cite{sanari2020modifying}.
Indeed, the shift current response is fast and would fully develop with a time scale $\mathcal{O}(10)$ - $\mathcal{O}(100)$ fs, determined by the material relaxation rate \cite{he2024ultrafast,sotome2019spectral}.
However, for the adopted resolution bandwidth of $10 \ \mathrm{GHz}$, we require relatively long pulses with durations of at least $100~\mathrm{ps}$ to collect sufficient data for spectral analysis. The simultaneous realization of the adopted long pulse durations and field strengths requires further technological development.
We note that shorter pulses may also be employed. Shorter pulse durations broaden the linewidth as $\sim 1/\tau_{\rm pulse}$ and thereby increase the thermal noise (since the resolution bandwidth $\Delta f_{\rm bin}$ is chosen to match the signal linewidth). Nevertheless, the resulting degradation in sensitivity remains weak ($g_{a\gamma\gamma}\propto \Delta f_{\rm bin}^{1/4}$, see Eq.~\eqref{eq:SNR}), provided that the linewidth remains within the material bandwidth. 
It is also worth noting that long and intense pulses may lead to potential material damage, the effects of which require careful experimental investigation and could potentially be mitigated by using shorter pulses.
Achieving a large beam size is also challenging. 
It must cover the sample size scale; otherwise, photothermoelectric currents or resistive losses become important and can be the background~\cite{osterhoudt2019colossal}.
subject for future work.

We note that the penetration depth of the sample is a crucial parameter that limits one dimension of the cross-section of a single sample. Therefore, to maximize the effective sample size while respecting the depth constraint, it may be necessary to employ a compact circuit composed of multiple thinly sliced samples. 

In this paper, we adopt the ferroelectric material and choose TaAs semimetal as a sample to discuss axion sensitivity with our method. The other material with a similar property such as the type-II Weyl semimetal $\rm TaIrTe_4$ \cite{Ma2018NonlinearPO} might be applicable as well. In addition to electronic excitations, the shift current response can also occur through quasiparticle excitations below the electronic band gap, including magnon \cite{morimoto2019shift,morimoto21} or phonon excitations \cite{morimoto24}, such as in perovskite manganite \cite{ogino2024terahertz} and $\rm BaTiO_3$ \cite{okamura2022photovoltaic}, respectively. 
These kinds of excitation have an energy scale of $0.1-1 \ \rm meV$ ($\sim 2.5-250$ GHz), and might be used to conveniently probe the smaller axion mass range compared to the case of using electronic excitation. Also, we might apply the same proposed method for probing hidden photons, another candidate of dark matter, by using an oscillating electric field to enhance the shift current induced by hidden photons. We left them as future work.

\ul{\textit{Acknowledgments:}}
We gratefully thank Masashi Kawasaki for discussions at an earlier stage of this work. Also, Koichi Hamaguchi and Takeo Moroi for reading and giving comments on the draft. This work is partially supported by JST SPRING, Grant No. JPMJSP2108, JSPS KAKENHI Grant No. 24KJ0713, JSPS KAKENHI Grant No. 25KJ0950, and JSPS KAKENHI Grant No. 23KJ0678.

\bibliography{refPRL}
\vspace{0.5em}
\onecolumngrid
\appendix
\section{Path integral formalism in quantum many-particle systems}\label{sec:PathIntegral}
In this section, we summarize the path integral formulation for quantum many-particle systems \cite{negele1988quantum}.
\subsection{Notation}
Let us begin with the second quantized Hamiltonian for electrons.
\begin{align}
    \widehat{H}_{0}
    =
    \sum_{a}
    \int\frac{d^{d}\bm{k}}{(2\pi)^{d}}
    \varepsilon_{\bm{k}a}c^{\dagger}_{\bm{k}a}c_{\bm{k}a}
    \label{eq:H_0}
\end{align}
where $\{a\}$ is the band label. The integration over $k$ is for the Brillouin zone. $\varepsilon_{\bm{k}a}$ is the eigenenergy and $c^{\dagger}_{\bm{k}a},\;c_{\bm{k}n}$ are the corresponding creation/annihilation operators satisfying the anti-commutation relation:
\begin{gather}
    \{c_{\bm{k}a},\;c^{\dagger}_{\bm{k}'a'}\}=(2\pi)^{d}\delta_{aa'}\delta^{d}(\bm{k}-\bm{k}'),\;\;
    \{c_{\bm{k}a},\;c_{\bm{k}'a'}\}=\{c^{\dagger}_{\bm{k}a},\;c^{\dagger}_{\bm{k}'a'}\}=0.
\end{gather}
Vacuum state $|0\rangle$ is defined by annihilation operators,
\begin{align}
    c_{\bm{k}a}|0\rangle \equiv0.
\end{align}
One electron state is defined by creation operators,
\begin{align}
    |\psi_{\bm{k}a}\rangle \equiv c^{\dagger}_{\bm{k}a}|0\rangle.
\end{align}
Using the position state $|\bm{r}\rangle$, the position representation of the wave function is defined by,
\begin{align}
    \psi_{\bm{k}a}(\bm{r})\equiv\langle \bm{r}|\psi_{\bm{k}a}\rangle.
\end{align}

\subsection{Fermion coherent states}
We define the coherent state of a fermion by the eigenstate of annihilation operators,
\begin{align}
    c_{\alpha}|\xi\rangle =\xi_{\alpha}|\xi\rangle
\end{align}
where $\xi_{\alpha}$ are Grassmann numbers\footnote{They obey Grassmann algebra:\\ $\{\xi_{\alpha},\;\xi_{\beta}\}=0,\;\;\{\xi_{\alpha},\;c_{\beta}^{(\dagger)}\}=0,\;\;(\xi_{\alpha}c_{\beta})^{\dagger}=c_{\beta}^{\dagger}\xi_{\alpha}^{*}$, etc.} and the label $\alpha$ represents general indices. Such a state can be realized by,
\begin{align}
    |\xi\rangle &=e^{-\sum_{\alpha}\xi_{\alpha}c_{\alpha}^{\dagger}}|0\rangle
=\prod_{\alpha}(1-\xi_{\alpha}c_{\alpha}^{\dagger})|0\rangle
\end{align}
If we take the different set of eigenvalues $\{\xi_{\alpha}'\}$, we can construct the different coherent state $|\xi'\rangle$. The overlap of two coherent states is
\begin{align}
    \langle \xi|\xi '\rangle =e^{\sum_{\alpha}\xi_{\alpha}^{*}\xi_{\alpha}'}
    \label{eq:overlap}
\end{align}
The closure relation is written by
\begin{align}
    \int\prod_{\alpha}d\xi_{\alpha}^{*}d\xi_{\alpha}
    e^{
    -\sum_{\alpha}\xi_{\alpha}^{*}\xi_{\alpha}
    }
    |\xi\rangle\langle\xi|
    =1
    \label{eq:closure}
\end{align}
For a complete set of states $\{|n\rangle\}$ in the Fock space, the trace of an operator $\widehat{A}$ can be written by,
\begin{align}
    \mathrm{Tr}\,\widehat{A}
    &=\sum_{n}
    \langle n|\widehat{A}|n\rangle
    =\int\prod_{\alpha}d\xi_{\alpha}^{*}d\xi_{\alpha}
    e^{
    -\sum_{\alpha}\xi_{\alpha}^{*}\xi_{\alpha}
    }
    \langle -\xi|\widehat{A}|\xi\rangle
    \label{eq:coherent_trace}
\end{align}

\subsection{Coherent state path integral}
Consider the Hamiltonian $\widehat{H}(c_{\alpha}^{\dagger},\;c_{\alpha})$ which is written in normal ordering. Then, the matrix element of the time evolution operator is

\begin{align}
    \langle\xi_{f}|
    e^{-\frac{i}{\hbar}\widehat{H}(t_{f}-t_{i})}
    |\xi_{i}\rangle
    &=
    \int_{\xi_{\alpha}(t_{i})=\xi_{\alpha,i}}^{\xi_{\alpha}(t_{f})=\xi_{\alpha,f}}
    \mathcal{D}\xi^{*}\mathcal{D}\xi\,
    e^{
    \sum_{\alpha}
    \xi_{\alpha}^{*}(t_{f})\xi_{\alpha}(t_{f})
    }
    \times
    e^{
    \frac{i}{\hbar}
    \int_{t_{i}}^{t_{f}}dt
    \left[
    \sum_{\alpha}
    i\hbar\xi_{\alpha}^{*}(t)\frac{\partial\xi_{\alpha}(t)}{\partial t}
    -H(\xi_{\alpha}^{*}(t),\xi_{\alpha}(t))
    \right]
    }
    \label{eq:path_integral_final}
    \\
    \mathcal{D}\xi^{*}\mathcal{D}\xi
    &=\lim_{M\to \infty}
    \prod_{k=1}^{M-1}
    \prod_{\alpha}
    d\xi_{\alpha,k}^{*}d\xi_{\alpha,k}.
\end{align}
The partition function is defined by,
\begin{align}
    Z\equiv
    \mathrm{Tr}\,
    e^{
    -\beta(\widehat{H}-\mu\widehat{N})
    }.
\end{align}
We expand it by coherent states using Eq.~\eqref{eq:coherent_trace},
\begin{align}
    Z=\int\prod_{\alpha}d\xi_{\alpha}^{*}d\xi_{\alpha}
    e^{
    -\sum_{\alpha}\xi_{\alpha}^{*}\xi_{\alpha}
    }
    \langle -\xi|
    e^{
    -\beta(\widehat{H}-\mu\widehat{N})
    }
    |\xi\rangle
    .
    \label{eq:Zmetric}
\end{align}
The bracket part is similar to $\langle\xi_{f}|e^{-\frac{i}{\hbar}\widehat{H}(t_{f}-t_{i})}|\xi_{i}\rangle$ in Eq.~\eqref{eq:path_integral_final}. We transform the preceding expression into the path integral form by performing the following three steps:
\begin{itemize}
    \item Analytic continuation: the domain of $t$ is extended from the real line to the complex plane.
    \item Redefinition of variables and fields: $t=-i\tau\;\;(t_{i}=-i\tau_{i},\;\;t_{f}=-i\tau_{f}),\;\;\xi (\tau)\equiv\xi (t)$.
    \item Boundary condition: $\tau_{i}=0,\;\;\tau_{f}=\beta,\;\;\xi_{i}=\xi,\;\;\xi_{f}=-\xi$.
\end{itemize}
Then, in $\hbar=1$ unit, Eq.~\eqref{eq:path_integral_final} becomes 
\begin{align}
    &\langle -\xi|
    e^{-i\widehat{H}(-i\beta -0)}
    |\xi\rangle\nonumber
    \\
    =&
    \int_{\xi_{\alpha}(\tau_{i})=\xi_{\alpha}}^{\xi_{\alpha}(\tau_{f})=-\xi_{\alpha}}
    \mathcal{D}\xi^{*}\mathcal{D}\xi\,
    e^{
    \sum_{\alpha}
    (-\xi_{\alpha}^{*})(-\xi_{\alpha})
    }
    \times
    e^{
    i
    \int_{\tau_{i}}^{\tau_{f}}(-id\tau)
    \left[
    \sum_{\alpha}
    i\xi_{\alpha}^{*}(\tau)\frac{\partial\xi_{\alpha}(\tau)}{-i\partial\tau}
    -H(\xi_{\alpha}^{*}(\tau),\xi_{\alpha}(\tau))
    \right]
    }\nonumber
    \\
    =&
    \int_{\xi_{\alpha}(\tau_{i})=\xi_{\alpha}}^{\xi_{\alpha}(\tau_{f})=-\xi_{\alpha}}
    \mathcal{D}\xi^{*}\mathcal{D}\xi\,
    e^{
    \sum_{\alpha}
    \xi_{\alpha}^{*}\xi_{\alpha}
    }
    \times
    e^{
    -
    \int_{\tau_{i}}^{\tau_{f}}d\tau
    \left[
    \sum_{\alpha}
    \xi_{\alpha}^{*}(\tau)\frac{\partial\xi_{\alpha}(\tau)}{\partial\tau}
    +H(\xi_{\alpha}^{*}(\tau),\xi_{\alpha}(\tau))
    \right]
    }.
\end{align}
Therefore, in the case of $\mu=0$, the partition function becomes
\begin{align}
    Z=
    \int_{\xi_{f}=-\xi_{i}}
    \mathcal{D}\xi^{*}\mathcal{D}\xi\,
    e^{
    -
    \int_{0}^{\beta}d\tau
    \left[
    \sum_{\alpha}
    \xi_{\alpha}^{*}(\tau)\frac{\partial\xi_{\alpha}(\tau)}{\partial\tau}
    +H(\xi_{\alpha}^{*}(\tau),\xi_{\alpha}(\tau))
    \right]
    },
\end{align}
where $\prod_{\alpha}d\xi_{\alpha}^{*}d\xi_{\alpha}$ in Eq.~\eqref{eq:Zmetric} is absorbed into $\mathcal{D}\xi^{*}\mathcal{D}\xi$.

\newpage
\section{Derivation of the interaction terms}\label{sec:interactionterm}
In this section, we derive the electromagnetic interaction terms as perturbations.
\subsection{Bloch state}
Assume a system with periodicity under translation  ($r_{i}\rightarrow r_{i}+L_{i}$). Following the Bloch theorem, the wave function can be expressed by
\begin{align}
    \psi_{\bm{k}a}(\bm{r})
    =
    e^{i\bm{k}\cdot\bm{r}}u_{\bm{k}a}(\bm{r})
\end{align}
where $u_{\bm{k}a}(\bm{r})$ is a periodic function:
\begin{align}
    u_{\bm{k}a}(\bm{r})
    =
    u_{\bm{k}a}(\bm{r}+m_{i}L_{i}),
    \label{eq:u_periodic}
\end{align}
with $m_i$ denoting integer.
It is an eigenfunction of a $k$-dependent Hamiltonian associated with the Hamiltonian given by Eq.~\eqref{eq:H_0}:
\begin{align}
    \widehat{H}_{0}(\bm{k})
    \equiv
    e^{-i\bm{k}\cdot\widehat{\bm{r}}}
    \widehat{H}_{0}
    e^{i\bm{k}\cdot\widehat{\bm{r}}}.
    \label{eq:H_k_mode}
\end{align}
The general one-particle state can be expanded by the set of eigenfunctions as
\begin{align}
    \Psi
    =\sum_{a}
    \int\frac{d^{d}\bm{k}}{(2\pi)^{d}}
    C_{\bm{k}a}
    |\psi_{\bm{k}a}\rangle.
\end{align}

\subsection{Covariant derivative}
Let us calculate the matrix elements of the position operator $\widehat{\bm{r}}$. The transition amplitude of the operator with general particle states is given by
\begin{align}
    \langle\Psi '|\widehat{\bm{r}}|\Psi\rangle
    =&
    \int d^{d}\bm{r}
    \langle\Psi '|\widehat{\bm{r}}|\bm{r}\rangle
    \langle\bm{r}|\Psi\rangle\nonumber\\
    =&
    \sum_{a',a}\int
    d^{d}\bm{r}
    \frac{d^{d}\bm{k}'}{(2\pi)^{d}}
    \frac{d^{d}\bm{k}}{(2\pi)^{d}}
    C_{\bm{k}'a'}^{*}
    ie^{-i\bm{k}'\cdot\bm{r}}
    u_{\bm{k}'a'}^{*}(\bm{r})
    e^{i\bm{k}\cdot\bm{r}}
    u_{\bm{k}a}(\bm{r})
    \left(
    \frac{\partial}{\partial\bm{k}}
    C_{\bm{k}a}
    \right)
    \label{eq:r_first}\\
    &+
    \sum_{a',a}\int
    d^{d}\bm{r}
    \frac{d^{d}\bm{k}'}{(2\pi)^{d}}
    \frac{d^{d}\bm{k}}{(2\pi)^{d}}
    C_{\bm{k}'a'}^{*}
    ie^{-i\bm{k}'\cdot\bm{r}}
    e^{i\bm{k}\cdot\bm{r}}
    u_{\bm{k}'a'}^{*}(\bm{r})
    \left(
    \frac{\partial}{\partial\bm{k}}
    u_{\bm{k}a}(\bm{r})
    \right)
    C_{\bm{k}a}.
    \label{eq:r_second}
\end{align}
The first term \eqref{eq:r_first} is
\begin{align}
    \int d^{d}\bm{r}
    e^{-i\bm{k}'\cdot\bm{r}}
    u_{\bm{k}'a'}^{*}(\bm{r})
    e^{i\bm{k}\cdot\bm{r}}
    u_{\bm{k}a}(\bm{r})
    =&
    \int d^{d}\bm{r}
    \psi_{\bm{k}'a'}^{*}(\bm{r})
    \psi_{\bm{k}a}(\bm{r})\nonumber\\
    =&
    (2\pi)^{d}\delta_{a'a}\delta(\bm{k}'-\bm{k}).
    \label{eq:r_first_result}
\end{align}
The second term \eqref{eq:r_second} is
\begin{align}
    \int d^{d}\bm{r}
    e^{-i\bm{k}'\cdot\bm{r}}
    e^{i\bm{k}\cdot\bm{r}}
    u_{\bm{k}'a'}^{*}(\bm{r})
    \frac{\partial}{\partial\bm{k}}
    u_{\bm{k}a}(\bm{r})
    =&
    \prod_{i=1}^{d}
    \int_{-\infty}^{\infty}dr_{i}
    e^{-i(k_{i}'-k_{i})r_{i}}
    u_{\bm{k}'a'}^{*}(\bm{r})
    \frac{\partial}{\partial\bm{k}}
    u_{\bm{k}a}(\bm{r})\nonumber\\
    =&
    \prod_{i=1}^{d}
    \sum_{m_{i}=-\infty}^{\infty}
    \int_{0}^{L_{i}}dr_{i}
    e^{-i(k_{i}'-k_{i})(r_{i}+m_{i}L_{i})}
    u_{\bm{k}'a'}^{*}(\bm{r})
    \frac{\partial}{\partial\bm{k}}
    u_{\bm{k}a}(\bm{r})\nonumber\\
    =&
    \prod_{i=1}^{d}
    \sum_{m_{i}=-\infty}^{\infty}
    e^{-i(k_{i}'-k_{i})m_{i}L_{i}}
    \int_{0}^{L_{i}}dr_{i}
    e^{-i(k_{i}'-k_{i})r_{i}}
    u_{\bm{k}'a'}^{*}(\bm{r})
    \frac{\partial}{\partial\bm{k}}
    u_{\bm{k}a}(\bm{r}),
\end{align}
where we divided the integration range by $L_{i}$ and used the periodic condition (Eq.~\eqref{eq:u_periodic}). 
We can derive the formula:
\begin{align}
    \sum_{m_{i}=-\infty}^{\infty}
    e^{-i(k_{i}'-k_{i})m_{i}L_{i}}
    =
    \frac{2\pi}{L_{i}}
    \delta(k_{i}'-k_{i}).
\end{align}
Therefore,
\begin{align}
    \int d^{d}\bm{r}
    e^{-i\bm{k}'\cdot\bm{r}}
    e^{i\bm{k}\cdot\bm{r}}
    u_{\bm{k}'a'}^{*}(\bm{r})
    \frac{\partial}{\partial\bm{k}}
    u_{\bm{k}a}(\bm{r})
    =&
    \prod_{i=1}^{d}
    \frac{2\pi}{L_{i}}
    \delta(k_{i}'-k_{i})
    \int_{0}^{L_{i}}dr_{i}
    e^{-i(k_{i}'-k_{i})r_{i}}\;
    u_{\bm{k}'a'}^{*}(\bm{r})
    \frac{\partial}{\partial\bm{k}}
    u_{\bm{k}a}(\bm{r})\nonumber\\
    =&
    \frac{(2\pi)^{d}}{V}
    \delta(\bm{k}'-\bm{k})
    \int_{V}d\bm{r}\;
    e^{-i(\bm{k}'-\bm{k})\cdot\bm{r}}\;
    u_{\bm{k}'a'}^{*}(\bm{r})
    \frac{\partial}{\partial\bm{k}}
    u_{\bm{k}a}(\bm{r})\nonumber\\
    =&
    \frac{(2\pi)^{d}}{V}
    \delta(\bm{k}'-\bm{k})
    \int_{V}d\bm{r}\;
    u_{\bm{k}'a'}^{*}(\bm{r})
    \frac{\partial}{\partial\bm{k}}
    u_{\bm{k}a}(\bm{r}),
    \label{eq:r_second_result}
\end{align}
where $V$ is the volume of a unit cell, $V=\prod_{i}L_{i}$. Substituting Eqs.~\eqref{eq:r_first_result} and  \eqref{eq:r_second_result} into \eqref{eq:r_first} and \eqref{eq:r_second}, we obtain
\begin{align}
    \langle\Psi '|\widehat{\bm{r}}|\Psi\rangle
    =&
    \sum_{a',a}\int
    \frac{d^{d}\bm{k}'}{(2\pi)^{d}}
    \frac{d^{d}\bm{k}}{(2\pi)^{d}}
    C_{\bm{k}'a'}^{*}
    i(2\pi)^{d}\delta_{a'a}\delta(\bm{k}'-\bm{k})
    \left(
    \frac{\partial}{\partial\bm{k}}
    C_{\bm{k}a}
    \right)\nonumber\\
    &+
    \sum_{a',a}\int
    \frac{d^{d}\bm{k}'}{(2\pi)^{d}}
    \frac{d^{d}\bm{k}}{(2\pi)^{d}}
    C_{\bm{k}'a'}^{*}
    i
    \frac{(2\pi)^{d}}{V}
    \delta(\bm{k}'-\bm{k})
    \int_{V}d\bm{r}\;
    u_{\bm{k}'a'}^{*}(\bm{r})
    \frac{\partial}{\partial\bm{k}}
    u_{\bm{k}a}(\bm{r})
    C_{\bm{k}a}\nonumber\\
    &=
    \sum_{a',a}\int
    \frac{d^{d}\bm{k}'}{(2\pi)^{d}}
    \frac{d^{d}\bm{k}}{(2\pi)^{d}}
    C_{\bm{k}'a'}^{*}\;
    i
    (2\pi)^{d}\delta(\bm{k}'-\bm{k})
    \left(
    \delta_{a'a}\frac{\partial}{\partial\bm{k}}
    -
    i\bm{\mathcal{A}}_{a'a}(\bm{k})
    \right)
    C_{\bm{k}a},
    \label{eq:r_expectation}
\end{align}
where we defined the Berry connection $\bm{\mathcal{A}}_{a'a}(\bm{k})$ as
\begin{align}
    \bm{\mathcal{A}}_{a'a}(\bm{k})=\frac{i}{V}
    \int_{V}d\bm{r}\;
    u_{\bm{k}'a'}^{*}(\bm{r})
    \frac{\partial}{\partial\bm{k}}
    u_{\bm{k}a}(\bm{r}).
\end{align}
When we define the covariant derivative operator:
\begin{align}
    \widehat{\bm{\mathcal{D}}}_{(\bm{k}'a')(\bm{k}a)}
    \equiv
    (2\pi)^{d}\delta(\bm{k}'-\bm{k})
    \left(
    \delta_{a'a}\frac{\partial}{\partial\bm{k}}
    -
    i\bm{\mathcal{A}}_{a'a}(\bm{k})
    \right),
\end{align}
by comparing the above two, we get the relation between $\widehat{\bm{r}}$ and $\widehat{\bm{\mathcal{D}}}$:
\begin{align}
    \widehat{\bm{r}}=i\widehat{\bm{\mathcal{D}}}.
    \label{eq:r_covariant}
\end{align}
Similarly, we can calculate the matrix element of a general operator $\widehat{\mathcal{O}}$ by
\begin{align}
    \langle\psi_{\bm{k}'a'}|\widehat{\mathcal{O}}|\psi_{\bm{k}a}\rangle
    &=
    (2\pi)^{d}\delta(\bm{k}'-\bm{k})
    \langle u_{\bm{k}'a'}|\widehat{\mathcal{O}}(\bm{k})|u_{\bm{k}a}\rangle_{V}
    \label{eq:matrix_element_general}\\
    \langle u_{\bm{k}'a'}|\widehat{\mathcal{O}}(\bm{k})|u_{\bm{k}a}\rangle_{V}
    &=\frac{1}{V}\int_{V}d\bm{r}\;
    u_{\bm{k}'a'}^{*}(\bm{r})\langle\bm{r}|
    \widehat{\mathcal{O}}(\bm{k})|u_{\bm{k}a}\rangle,
\end{align}
where $\widehat{\mathcal{O}}(\bm{k})\equiv e^{-i\bm{k}\cdot\widehat{\bm{r}}}\widehat{\mathcal{O}}e^{i\bm{k}\cdot\widehat{\bm{r}}}$ is the $k$-dependent operator.
\subsection{Electromagnetic interactions}
The electromagnetic perturbation is realized by gauge principle replacing $\bm{k}\rightarrow\bm{k}-q\bm{A}(t)$ where the vector potential $A(t)$ is chosen so that $\bm{E}(t)=-\partial_{t}\bm{A}(t)$.
\begin{align}
    \widehat{H}_{A}(\bm{k},t)
    =
    \widehat{H}_{0}
    \left(
    \bm{k}-q\bm{A}(t)
    \right).
\end{align}
Using Eqs.~\eqref{eq:H_k_mode} and \eqref{eq:r_covariant} and the Baker-Campbell-Housdorff formula,
\begin{align}
    e^{\widehat{A}}\widehat{B}e^{-\widehat{A}}
    =
    \widehat{B}
    +
    \left[
    \widehat{A},\widehat{B}
    \right]
    +
    \frac{1}{2}
    \left[
    \widehat{A},
     \left[
     \widehat{A},\widehat{B}
     \right]
    \right]
    +
    \cdots
    ,
\end{align}
the Hamiltonian becomes 
\begin{align}
    \widehat{H}_{A}(\bm{k},t)
    =&\;
    e^{
    -i
    (\bm{k}-q\bm{A})
    \cdot\widehat{\bm{r}}
    }\;
    \widehat{H}_{0}\;
    e^{
    i
    (\bm{k}-q\bm{A})
    \cdot\widehat{\bm{r}}
    }\nonumber\\
    =&\;
    e^{
    q\bm{A}
    \cdot
    \widehat{\bm{\mathcal{D}}}
    }\;
    \widehat{H}_{0}(\bm{k})\;
    e^{
    -q\bm{A}
    \cdot
    \widehat{\bm{\mathcal{D}}}
    }\nonumber\\
    =&\;
    \widehat{H}_{0}(\bm{k})
    +
    \left[
    q\bm{A}\cdot\widehat{\bm{\mathcal{D}}},
    \widehat{H}_{0}(\bm{k})
    \right]
    +
    \frac{1}{2}
    \left[
    q\bm{A}\cdot\widehat{\bm{\mathcal{D}}},
     \left[
     q\bm{A}\cdot\widehat{\bm{\mathcal{D}}},
     \widehat{H}_{0}(\bm{k})
     \right]
    \right]
    +
    \cdots\nonumber\\
    =&\;
    \widehat{H}_{0}(\bm{k})
    +
    \sum_{n=1}^{\infty}
    \frac{1}{n!}
    \left[
    \prod_{k=1}^{n}
    qA^{\alpha_{k}}
    \widehat{\mathcal{D}}^{\alpha_{k}}
    \right]
    \widehat{H}_{0}(\bm{k})\nonumber\\
    =&\;
    \widehat{H}_{0}(\bm{k})
    +
    \sum_{n=1}^{\infty}
    \frac{1}{n!}
    \prod_{k=1}^{n}
    qA^{\alpha_{k}}
    \widehat{h}^{\alpha_{1}...\alpha_{n}}(\bm{k}),
\end{align}
where $\alpha_{k}$ is a spatial index with an implicit sum.
Note that we use the following shorthands:
\begin{align}
    \widehat{\bm{\mathcal{D}}}[\widehat{\mathcal{O}}]
    \equiv &\;
    [\widehat{\bm{\mathcal{D}}},\;\widehat{\mathcal{O}}]\\
    \widehat{h}^{\alpha_{1}...\alpha_{n}}
    \equiv &\;
    \widehat{\mathcal{D}}^{\alpha_{1}}
    \cdots\widehat{\mathcal{D}}^{\alpha_{n}}[\widehat{H}_{0}].
\end{align}
Finally, we go back to the $k$-independent form
\footnote{
$\widehat{\bm{\mathcal{D}}}$ and $e^{\pm i\bm{k}\cdot\widehat{\bm{r}}}$ are commutable since $\widehat{\bm{r}}=i\widehat{\bm{\mathcal{D}}}$. Therefore, we can use $\widehat{\bm{\mathcal{D}}}=e^{i\bm{k}\cdot\widehat{\bm{r}}}\widehat{\bm{\mathcal{D}}}e^{-i\bm{k}\cdot\widehat{\bm{r}}}$ to derive \eqref{eq:Hamiltonian_Taylor2}.
}
,
\begin{align}
    \widehat{H}_{A}(t)
    \equiv &\;
    e^{i\bm{k}\cdot\widehat{\bm{r}}}
    \widehat{H}_{A}(\bm{k},t)
    e^{-i\bm{k}\cdot\widehat{\bm{r}}}\nonumber\\
    =&\;
    \widehat{H}_{0}
    +
    \sum_{n=1}^{\infty}
    \frac{1}{n!}
    \prod_{k=1}^{n}
    qA^{\alpha_{k}}
    \widehat{h}^{\alpha_{1}...\alpha_{n}}.
    \label{eq:Hamiltonian_Taylor2}
\end{align}
The second term of \eqref{eq:Hamiltonian_Taylor2} represents the perturbation due to the external field ($\equiv V_{E}(t)$).

We perform the Fourier transformation:
\begin{align}
    \bm{A}(t)
    =&
    \int d\omega
    e^{-i(\omega+i\gamma)t}
    \bm{A}(\omega)
    \label{eq:delta_factor}
    \\
    \bm{E}(\omega)
    =&\;
    i\omega\bm{A}(\omega),
\end{align}
where $\gamma$ is a small and positive real number. It satisfies the assumption that the initial state is in equilibrium $A(t)|_{t\rightarrow -\infty}=0$. We are interested in the evolution from $t=-\infty \rightarrow t \sim 0$ to find the steady-state solution. To avoid notational complexity, we omit $\gamma$ in the following calculations and reintroduce it later. The perturbation becomes
\begin{align}
    V_{E}(t)
    &=
    \sum_{m=1}^{\infty}
    \frac{1}{m!}
    \prod_{j=1}^{m}
    \int d\omega_{j}
    e^{-i\omega_{j} t}
    \left(
    \frac{-iq}{\omega_{j}}
    \right)
    E^{\alpha_{j}}(\omega_{j})
    \widehat{h}^{\alpha_{1}...\alpha_{m}}\nonumber\\
    &=
    \sum_{m=1}^{\infty}
    \frac{1}{m!}
    \prod_{j=1}^{m}
    \int d\omega_{j}
    e^{-i\omega_{j} t}
    \left(
    \frac{ie}{\omega_{j}}
    \right)
    E^{\alpha_{j}}(\omega_{j})
    \widehat{h}^{\alpha_{1}...\alpha_{m}}, \label{eq:appendixVE}
\end{align}
where $q=-e,\;(e>0)$ for the electron case.
Using Eq.~\eqref{eq:matrix_element_general}, $\widehat{h}$ can be expanded by its matrix elements $h_{\bm{k}aa'}\equiv\langle\psi_{\bm{k}a}|\widehat{h}|\psi_{\bm{k}a'}\rangle$ which means
\begin{align}
    \widehat{h}^{\alpha_{1}...\alpha_{n}}
    =
    \sum_{aa'}
    \int\frac{d^{d}\bm{k}}{(2\pi)^{d}}
    c^{\dagger}_{\bm{k}a}
    h_{\bm{k}aa'}^{\alpha_{1}...\alpha_{n}}
    c_{\bm{k}a'}.
\end{align}

\newpage
\section{General calculation of the optical response}\label{sec:opticalresponse}
In this section, we calculate the optical response up to the second order.
\subsection{Partition function}
Consider the path-integral form of the partition function:
\begin{align}
    Z
    =
    \int_{\xi_{f}=-\xi_{i}}
    \mathcal{D}\xi^{*}\mathcal{D}\xi\,
    e^{-S}
\end{align}
\begin{align}
    S
    =
    \int_{0}^{\beta}d\tau
    \sum_{aa'}
    \int\frac{d^{d}\bm{k}}{(2\pi)^{d}}
    \left[
    \delta_{aa'}
    \xi_{\bm{k}a}^{*}(\tau)
    \frac{\partial\xi_{\bm{k}a'}(\tau)}{\partial\tau}
    +H_{0}(\xi_{\bm{k}a}^{*}(\tau),\xi_{\bm{k}a'}(\tau))
    +V_{E}(\xi_{\bm{k}a}^{*}(\tau),\xi_{\bm{k}a'}(\tau))
    \right]
\end{align}
\begin{align}
    H_{0}(\xi_{\bm{k}a}^{*}(\tau),\xi_{\bm{k}a'}(\tau))
    =&\,
    \delta_{aa'}
    \xi_{\bm{k}a}^{*}(\tau)
    \varepsilon_{\bm{k}a'}
    \xi_{\bm{k}a'}(\tau)
    \\
    V_{E}(\xi_{\bm{k}a}^{*}(\tau),\xi_{\bm{k}a'}(\tau))
    =&\,
    \sum_{m=1}^{\infty}
    \frac{1}{m!}
    \prod_{j=1}^{m}
    i\int d\omega_{n;j}
    e^{-i\omega_{n;j} \tau}
    \left(
    \frac{e}{\omega_{n;j}}
    \right)
    E^{\alpha_{j}}(\omega_{n;j})
    \xi_{\bm{k}a}^{*}(\tau)
    h_{\bm{k}aa'}^{\alpha_{1}...\alpha_{m}}
    \xi_{\bm{k}a'}(\tau),
\end{align}
where\footnote{Since $\omega_{n;j}$ is discrete, $\int d\omega_{n;j}$ actually means $\sum_{\omega_{n;j}}$.\\For the convenience, we keep to white it as $\int d\omega_{n;j}$.}
\begin{align}
    \tau=it,\;\;\omega_{n;j}=-i\omega_{j},\;\;E^{\alpha_{j}}(\omega_{n;j})\equiv E^{\alpha_{j}}(\omega_{j}),\;\;V_{E}(\tau)\equiv\; V_{E}(t).
\end{align}
We perform the Fourier transformation
\footnote{
The inverse transformation is:\\$\xi_{\bm{k}a}(\omega_{n})=\int_{0}^{\beta}d\tau\xi_{\bm{k}a}(\tau)e^{i\omega_{n}\tau},\;\;\xi^{*}_{\bm{k}a}(\omega_{n})=\int_{0}^{\beta}d\tau\xi^{*}_{\bm{k}a}(\tau)e^{-i\omega_{n}\tau}$.
}
:
\begin{align}
    \xi_{\bm{k}a}(\tau)
    =
    \frac{1}{\beta}\sum_{n}
    \xi_{\bm{k}a}(\omega_{n})e^{-i\omega_{n}\tau}
    ,\;\;\;\;
    \xi^{*}_{\bm{k}a}(\tau)
    =
    \frac{1}{\beta}\sum_{n}
    \xi^{*}_{\bm{k}a}(\omega_{n})e^{i\omega_{n}\tau},
    \label{eq:xi_Fourier}
\end{align}
where $\omega_{n}$ is called Matsubara frequency ($\omega_{n}=(2n+1)\pi/\beta$)\footnote{This discreteness is from the boundary condition $\xi_{f}=-\xi_{i}$.}.
Then, the nonperturbative action $S_{0}$ becomes
\begin{align}
    S_{0}
    =&
    \int_{0}^{\beta}d\tau
    \sum_{a}
    \int\frac{d^{d}\bm{k}}{(2\pi)^{d}}
    \left[
    \xi_{\bm{k}a}^{*}(\tau)
    \frac{\partial\xi_{\bm{k}a}(\tau)}{\partial\tau}
    +
    \xi_{\bm{k}a}^{*}(\tau)
    \varepsilon_{\bm{k}a}
    \xi_{\bm{k}a}(\tau)
    \right]\nonumber
    \\
    =&
    \int_{0}^{\beta}d\tau
    \sum_{a}
    \int\frac{d^{d}\bm{k}}{(2\pi)^{d}}
    \frac{1}{\beta^{2}}\sum_{n,n'}
    e^{i(\omega_{n}-\omega_{n'})\tau}
    \left[
    -\xi_{\bm{k}a}^{*}(\omega_{n})i\omega_{n'}\xi_{\bm{k}a}(\omega_{n'})
    +\xi_{\bm{k}a}^{*}(\omega_{n})\varepsilon_{\bm{k}a}\xi_{\bm{k}a}(\omega_{n'})
    \right].
\end{align}
We can calculate $\tau$ integration with
\begin{align}
    \int_{0}^{\beta}d\tau
    e^{i(\omega_{n}-\omega_{n'})\tau}
    =
    \beta\delta_{nn'}.
\end{align}
The Fourier-transformed action is
\begin{align}
    S_{0}
    =&
    \sum_{a}
    \int\frac{d^{d}\bm{k}}{(2\pi)^{d}}
    \frac{1}{\beta^{2}}\sum_{n,n'}
    \beta\delta_{nn'}
    \left[
    -\xi_{\bm{k}a}^{*}(\omega_{n})
    i\omega_{n'}
    \xi_{\bm{k}a}(\omega_{n'})
    +
    \xi_{\bm{k}a}^{*}(\omega_{n})
    \varepsilon_{\bm{k}a}
    \xi_{\bm{k}a}(\omega_{n'})
    \right]\nonumber
    \\
    =&
    \sum_{a}
    \int\frac{d^{d}\bm{k}}{(2\pi)^{d}}
    \frac{1}{\beta}
    \sum_{n}
    \left[
    -\xi_{\bm{k}a}^{*}(\omega_{n})
    i\omega_{n}
    \xi_{\bm{k}a}(\omega_{n})
    +
    \xi_{\bm{k}a}^{*}(\omega_{n})
    \varepsilon_{\bm{k}a}
    \xi_{\bm{k}a}(\omega_{n})
    \right],
\end{align}
and the nonperturbative partition function $Z_{0}$ becomes
\begin{align}
    Z_{0}
    =&
    |J|\int\mathcal{D}\xi^{*}\mathcal{D}\xi\,
    \prod_{n,\bm{k},a}
    e^{
    -\xi^{*}_{\bm{k}a}(\omega_{n})
    (-i\omega_{n}+\varepsilon_{\bm{k}a})
    \xi_{\bm{k}a}(\omega_{n})
    }\nonumber
    \\
    =&
    |J|\prod_{n,\bm{k},a}
    (-1)\frac{1}{\beta}(i\omega_{n}-\varepsilon_{\bm{k}a}),
\end{align}
where $|J|$ is the Jacobian from Eq.~\eqref{eq:xi_Fourier} and we used the Gaussian integral for Grassmann variables:
\begin{align}
    \int d\xi^{*}d\xi\,
    e^{-\xi^{*}M\xi}=\mathrm{det}M.
\end{align}
\subsection{Green function}
We introduce the Green function
\begin{align}
    G_{\bm{k}aa'}(\tau,\tau')
    &\equiv
    -\langle\mathrm{T}c_{\bm{k}a}(\tau)c^{\dagger}_{\bm{k}a'}(\tau')\rangle
    =
    -\mathrm{Tr}
    \left(
    e^{-\beta\widehat{H}}c_{\bm{k}a}(\tau)c^{\dagger}_{\bm{k}a'}(\tau')
    \right).
\end{align}
Applying the Heisenberg picture $c_{\bm{k}a}(\tau)=e^{\tau\widehat{H}}c_{\bm{k}a}(0)e^{-\tau\widehat{H}},\;\;c^{\dagger}_{\bm{k}a}(\tau)=e^{\tau\widehat{H}}c^{\dagger}_{\bm{k}a}(0)e^{-\tau\widehat{H}}$,
\begin{align}
    G_{\bm{k}aa'}(\tau,\tau')
    =&
    -\mathrm{Tr}
    \left(
    e^{-\beta\widehat{H}}
    e^{\tau\widehat{H}}c_{\bm{k}a}(0)e^{-\tau\widehat{H}}
    e^{\tau'\widehat{H}}c^{\dagger}_{\bm{k}a'}(0)e^{-\tau'\widehat{H}}
    \right)\nonumber
    \\
    =&
    -\mathrm{Tr}
    \left(
    e^{-\beta\widehat{H}}
    e^{(\tau-\tau')\widehat{H}}c_{\bm{k}a}(0)e^{-(\tau-\tau')\widehat{H}}
    c^{\dagger}_{\bm{k}a'}(0)
    \right)\nonumber
    \\
    =&
    G_{\bm{k}aa'}(\tau-\tau',0),
\end{align}
where we used the trace identity $\mathrm{Tr}(ABC)=\mathrm{Tr}(CAB)$. Therefore, the Green function depends only on $\tau-\tau'$ and we can replace $\tau-\tau'$ to $\tau$ without loss of  generality. Here is the new convention:
\begin{align}
    G_{\bm{k}aa'}(\tau)
    \equiv
    -\langle\mathrm{T}c_{\bm{k}a}(\tau)c^{\dagger}_{\bm{k}a'}(0)\rangle.
\end{align}
The path integral form for each convention is
\begin{align}
    G_{\bm{k}aa'}(\tau,\tau')
    =&
    -\frac{1}{Z_{0}}
    \int\mathcal{D}\xi^{*}\mathcal{D}\xi\,
    \xi_{\bm{k}a}(\tau)\xi^{*}_{\bm{k}a'}(\tau')e^{-S_{0}},
    \label{eq:gleen_def}
    \\
    G_{\bm{k}aa'}(\tau)
    =&
    -\frac{1}{Z_{0}}
    \int\mathcal{D}\xi^{*}\mathcal{D}\xi\,
    \xi_{\bm{k}a}(\tau)\xi^{*}_{\bm{k}a'}(0)e^{-S_{0}}.
\end{align}
Let us derive its Fourier transformation. Applying Eq.~\eqref{eq:xi_Fourier}, we obtain
\begin{align}
    G_{\bm{k}aa'}(\tau)
    =
    -\frac{1}{Z_{0}}|J|
    \int\mathcal{D}\xi^{*}\mathcal{D}\xi\,
    \frac{1}{\beta^{2}}\sum_{n,n'}
    \xi_{\bm{k}a}(\omega_{n})e^{-i\omega_{n}\tau}\xi^{*}_{\bm{k}a'}(\omega_{n'})e^{-S_{0}}.
\end{align}
This Grassmann integral has a nonzero value only for $(n,a)=(n',a')$. Thus,
\begin{align}
    G_{\bm{k}aa'}(\tau)
    =&
    -\frac{1}{Z_{0}}|J|
    \int\mathcal{D}\xi^{*}\mathcal{D}\xi\,
    \frac{1}{\beta^{2}}\sum_{n,n'}
    \delta_{nn'}\delta_{aa'}
    \xi_{\bm{k}a}(\omega_{n})e^{-i\omega_{n}\tau}\xi^{*}_{\bm{k}a'}(\omega_{n'})e^{-S_{0}}\nonumber
    \\
    =&
    \frac{1}{\beta}\sum_{n}
    \left(
    -\frac{1}{Z_{0}}|J|\delta_{aa'}
    \int\mathcal{D}\xi^{*}\mathcal{D}\xi\,
    \frac{1}{\beta}
    \xi_{\bm{k}a}(\omega_{n})\xi^{*}_{\bm{k}a'}(\omega_{n})e^{-S_{0}}
    \right)
    e^{-i\omega_{n}\tau}.
    \label{eq:gleen_fourier_relation}
\end{align}
Then, we can define and calculate the Fourier transformation as
\begin{align}
    G_{\bm{k}aa'}(\omega_{n})
    \equiv&
    -\frac{1}{Z_{0}}|J|\delta_{aa'}
    \int\mathcal{D}\xi^{*}\mathcal{D}\xi\,
    \frac{1}{\beta}
    \xi_{\bm{k}a}(\omega_{n})\xi^{*}_{\bm{k}a'}(\omega_{n})e^{-S_{0}}\nonumber
    \\
    =&
    -\delta_{aa'}
    \frac{1}{\beta}
    \frac{
    |J|\prod_{(m,\bm{k}',a'')\neq(n,\bm{k},a)}
    (-1)(1/\beta)(i\omega_{m}-\varepsilon_{\bm{k'}a''})
    }
    {
    |J|\prod_{(m,\bm{k}',a'')}
    (-1)(1/\beta)(i\omega_{m}-\varepsilon_{\bm{k'}a''})
    }\nonumber
    \\
    =&
    \delta_{aa'}\frac{1}{i\omega_{n}-\varepsilon_{\bm{k}a}},
    \label{eq:gleen_fourier_final}
\end{align}
where we used the formula
\begin{align}
    \int d\xi^{*}d\xi\,
    \xi\xi^{*}e^{-\xi^{*}M\xi}
    =
    \int d\xi^{*}d\xi\,
    \xi\xi^{*}(1-\xi^{*}M\xi)
    =
    1.
\end{align}
Substituting Eq.~\eqref{eq:gleen_fourier_final} into Eq.~\eqref{eq:gleen_fourier_relation}, we obtain the inverse Fourier transformation as
\begin{align}
    G_{\bm{k}aa'}(\tau)
    =&
    \frac{1}{\beta}\sum_{n}
    \left(
    \delta_{aa'}\frac{1}{i\omega_{n}-\varepsilon_{\bm{k}a}}
    \right)
    e^{-i\omega_{n}\tau}
    =
    \delta_{aa'}
    \frac{1}{\beta}
    \sum_{n}
    \frac{e^{-i\omega_{n}\tau}}{i\omega_{n}-\varepsilon_{\bm{k}a}}.
\end{align}

\subsection{Source fields}
We introduce source fields $\{\overline{\alpha}_{\bm{k}a}(\tau),\;\gamma_{\bm{k}a}(\tau)\}$ for $\{\xi_{\bm{k}a}(\tau),\;\xi_{\bm{k}a}^{*}(\tau)\}$. Then, the nonperturbative action $S_{0,\overline{\alpha},\gamma}$ is
\begin{align}
    S_{0,\overline{\alpha},\gamma}
    =
    \int_{0}^{\beta}d\tau
    \sum_{a}
    \int\frac{d^{d}\bm{k}}{(2\pi)^{d}}
    \left[
    \xi_{\bm{k}a}^{*}(\tau)\frac{\partial\xi_{\bm{k}a}(\tau)}{\partial\tau}
    +\xi_{\bm{k}a}^{*}(\tau)\varepsilon_{\bm{k}a}\xi_{\bm{k}a}(\tau)
    +\overline{\alpha}_{\bm{k}a}(\tau)\xi_{\bm{k}a}(\tau)
    +\xi_{\bm{k}a}^{*}(\tau)\gamma_{\bm{k}a}(\tau)
    \right].
\end{align}
We perform the Fourier transformation:
\begin{align}
    \gamma_{\bm{k}}(\tau)=\frac{1}{\beta}\sum_{n}\gamma_{\bm{k}}(\omega_{n})e^{-i\omega_{n}\tau}
    ,\;\;\;\;
    \overline{\alpha}_{\bm{k}}(\tau)=\frac{1}{\beta}\sum_{n}\overline{\alpha}_{\bm{k}}(\omega_{n})e^{i\omega_{n}\tau}
    .
\end{align}
The action becomes
\begin{align}
    S_{0,\overline{\alpha},\gamma}
    =
    \sum_{a}
    \int\frac{d^{d}\bm{k}}{(2\pi)^{d}}
    \frac{1}{\beta}\sum_{n}
    \left[
    \xi_{\bm{k}a}^{*}(\omega_{n})
    (-i\omega_{n}+\varepsilon_{\bm{k}a})
    \xi_{\bm{k}a}(\omega_{n})
    +\overline{\alpha}_{\bm{k}a}(\omega_{n})\xi_{\bm{k}a}(\omega_{n})
    +\xi_{\bm{k}a}^{*}(\omega_{n})\gamma_{\bm{k}a}(\omega_{n})
    \right].
\end{align}
We calculate the inside part as
\begin{align}
    &\xi_{\bm{k}a}^{*}(\omega_{n})(-i\omega_{n}+\varepsilon_{\bm{k}a})\xi_{\bm{k}a}(\omega_{n})
    +\overline{\alpha}_{\bm{k}a}(\omega_{n})\xi_{\bm{k}a}(\omega_{n})
    +\xi_{\bm{k}a}^{*}(\omega_{n})\gamma_{\bm{k}a}(\omega_{n})\nonumber
    \\
    =&
    (-i\omega_{n}+\varepsilon_{\bm{k}a})
    \left[
    \xi_{\bm{k}a}^{*}(\omega_{n})\xi_{\bm{k}a}(\omega_{n})
    +
    \frac{\overline{\alpha}_{\bm{k}a}(\omega_{n})}{-i\omega_{n}+\varepsilon_{\bm{k}a}}
    \xi_{\bm{k}a}(\omega_{n})
    +
    \xi_{\bm{k}a}^{*}(\omega_{n})
    \frac{\gamma_{\bm{k}a}(\omega_{n})}{-i\omega_{n}+\varepsilon_{\bm{k}a}}
    \right]\nonumber
    \\
    =&
    (-i\omega_{n}+\varepsilon_{\bm{k}a})
    \left[
        \left(
        \xi_{\bm{k}a}^{*}(\omega_{n})
        +
        \frac{\overline{\alpha}_{\bm{k}a}(\omega_{n})}{-i\omega_{n}+\varepsilon_{\bm{k}a}}
        \right)
        \left(
        \xi_{\bm{k}a}(\omega_{n})
        +
        \frac{\gamma_{\bm{k}a}(\omega_{n})}{-i\omega_{n}+\varepsilon_{\bm{k}a}}
        \right)
        -
        \frac{\overline{\alpha}_{\bm{k}a}(\omega_{n})}{-i\omega_{n}+\varepsilon_{\bm{k}a}}
        \frac{\gamma_{\bm{k}a}(\omega_{n})}{-i\omega_{n}+\varepsilon_{\bm{k}a}}
    \right].
\end{align}
We shift the fields:
\begin{align}
    \chi_{\bm{k}a}^{*}(\omega_{n})
    =
    \xi_{\bm{k}a}^{*}(\omega_{n})
    +
    \frac{\overline{\alpha}_{\bm{k}a}(\omega_{n})}{-i\omega_{n}+\varepsilon_{\bm{k}a}}
    ,\;\;\;\;
    \chi_{\bm{k}a}(\omega_{n})
    =
    \xi_{\bm{k}a}(\omega_{n})
    +
    \frac{\gamma_{\bm{k}a}(\omega_{n})}{-i\omega_{n}+\varepsilon_{\bm{k}a}}.
\end{align}
Then,
\begin{align}
    Z_{0,\overline{\alpha},\gamma}
    =&
    \int
    \mathcal{D}\chi^{*}\mathcal{D}\chi\,
    \mathrm{exp}
    \left[
    \sum_{a}
    \int\frac{d^{d}\bm{k}}{(2\pi)^{d}}
    \frac{1}{\beta}
    \sum_{n}
    -\chi_{\bm{k}a}^{*}(\omega_{n})(-i\omega_{n}+\varepsilon_{\bm{k}a})\chi_{\bm{k}a}(\omega_{n})
    +
    \frac{\overline{\alpha}_{\bm{k}a}(\omega_{n})\gamma_{\bm{k}a}(\omega_{n})}{-i\omega_{n}+\varepsilon_{\bm{k}a}}
    \right]\nonumber
    \\
    =&
    Z_{0}\,
    \mathrm{exp}
    \left[
    \sum_{a}
    \int\frac{d^{d}\bm{k}}{(2\pi)^{d}}
    \frac{1}{\beta}
    \sum_{n}
    \overline{\alpha}_{\bm{k}a}(\omega_{n})
    \frac{1}{-i\omega_{n}+\varepsilon_{\bm{k}a}}
    \gamma_{\bm{k}a}(\omega_{n})
    \right]\nonumber
    \\
    =&
    Z_{0}\,
    \mathrm{exp}
    \left[
    \sum_{aa'}
    \int\frac{d^{d}\bm{k}}{(2\pi)^{d}}d\tau d\tau'
    \overline{\alpha}_{\bm{k}a}(\tau)
    \left(
    -
    G_{\bm{k}aa'}(\tau-\tau')
    \right)
    \gamma_{\bm{k}a'}(\tau')
    \right].
    \label{eq:-g}
\end{align}
We can derive that\footnote{The minus sign in Eq.~\eqref{eq:-g} is canceled by the anticommutation of $\delta\gamma$ and $\overline{\alpha}$ in the first derivative.}
\begin{align}
    G_{\bm{k}aa'}(\tau-\tau')
    =
    \left.
    \frac{1}{Z_{0}}
    \frac{\delta}{\delta\overline{\alpha}_{\bm{k}a}(\tau)}
    \frac{\delta}{\delta\gamma_{\bm{k}a'}(\tau')}
    Z_{0,\overline{\alpha},\gamma}
    \right|_{\overline{\alpha}=\gamma=0}.
    \label{eq:gleen_partial}
\end{align}
We have the formula:
\begin{align}
    \left.
    \frac{\delta}{\delta\overline{\alpha}_{\bm{k}a}(\tau)}
    Z_{0,\overline{\alpha},\gamma}
    \right|_{\overline{\alpha}=\gamma=0}
    =
    \int\mathcal{D}\xi^{*}\mathcal{D}\xi\,
    \xi_{\bm{k}a}(\tau)e^{-S_{0}},
\end{align}
\begin{align}
    \left.
    \frac{-\delta}{\delta\gamma_{\bm{k}a'}(\tau')}
    Z_{0,\overline{\alpha},\gamma}
    \right|_{\overline{\alpha}=\gamma=0}
    =
    \int\mathcal{D}\xi^{*}\mathcal{D}\xi\,
    \xi^{*}_{\bm{k}a'}(\tau')e^{-S_{0}}.
\end{align}
Therefore, with the interaction, the partition function becomes
\begin{align}
    Z
    =&\,
    \left.
    \mathrm{exp}
    \left[-
    \int_{0}^{\beta} d\tau\, 
    \sum_{aa'}
    \int\frac{d^{d}\bm{k}}{(2\pi)^{d}}
    V_{E}
        \left(
        \frac{\delta}{\delta\overline{\alpha}_{\bm{k}a}(\tau)},
        \frac{-\delta}{\delta\gamma_{\bm{k}a'}(\tau)}
        \right)
    \right]
    Z_{0,\overline{\alpha},\gamma}
    \right|_{\overline{\alpha}=\gamma=0}\nonumber
    \\
    =&\,
    \mathrm{exp}
    \left[-
    \int_{0}^{\beta} d\tau\, 
    \sum_{aa'}
    \int\frac{d^{d}\bm{k}}{(2\pi)^{d}}
    \sum_{m=1}^{\infty}
    \frac{1}{m!}
    \prod_{j=1}^{m}
    i\int d\omega_{n;j}
    e^{-i\omega_{n;j} \tau}
    \right.\nonumber
    \\
    &\;\;\;\;\;\;\;\;\;\;\;\;\;\;\;\;\;\;\;\;\times
    \left.
    \left(
    \frac{e}{\omega_{n;j}}
    \right)
    E^{\alpha_{k}}(\omega_{n;j})
    \frac{-\delta}{\delta\gamma_{\bm{k}a}(\tau)}
    h_{\bm{k}aa'}^{\alpha_{1}...\alpha_{m}}
    \frac{\delta}{\delta\overline{\alpha}_{\bm{k}a'}(\tau)}
    \right]
    \left.
    Z_{0,\overline{\alpha},\gamma}
    \right|_{\overline{\alpha}=\gamma=0}.
\end{align}

\subsection{The expectation value of the current}
The expectation value of the output current is
\begin{align}
    \langle\widehat{J}^{\mu}\rangle (\tau)
    =
    \frac{1}{Z}
    \int\mathcal{D}\xi^{*}\mathcal{D}\xi\,
    \left[
    ev_{E}^{\mu}(\tau)e^{-S}
    \right],
    \label{eq:path_integral_current}
\end{align}
where $\widehat{v}_{E}^{\mu}(\tau)$ is the velocity operator obtained by the analytic continuation $\widehat{v}_{E}^{\mu}(\tau)\equiv\widehat{v}_{E}^{\mu}(t)$, with $\widehat{v}_{E}^{\mu}(t)\equiv\,\dot{\widehat{r}}^{\mu}=
    -i[\widehat{r}^{\mu},\,\widehat{H}_{0}+\widehat{V}_{E}(t)]$ governed by the Heisenberg equation of motion:   
\begin{align}
    \widehat{v}_{E}^{\mu}(\tau)
    \equiv&
    \,
    \widehat{\mathcal{D}}^{\mu}
    [
    \widehat{H}_{0}+\widehat{V}_{E}(\tau)
    ]\nonumber
    \\
    =&\,
    \widehat{h}^{\mu}
    +
    \sum_{m=1}^{\infty}
    \frac{1}{m!}
    \prod_{j=1}^{m}
    i\int d\omega_{n;j}
    e^{-i\omega_{n;j} \tau}
    \left(
    \frac{e}{\omega_{n;j}}
    \right)
    E^{\alpha_{j}}(\omega_{n;j})
    \widehat{h}^{\mu\alpha_{1}...\alpha_{m}}\nonumber
    \\
    =&
    \sum_{aa'}
    \int\frac{d^{d}\bm{k}}{(2\pi)^{d}}
    \left\{
    c^{\dagger}_{\bm{k}a}
    h_{\bm{k}aa'}^{\mu}
    c_{\bm{k}a'}
    +
    \sum_{m=1}^{\infty}
    \frac{1}{m!}
    \prod_{j=1}^{m}
    i\int d\omega_{n;j}
    e^{-i\omega_{n;j} \tau}
    \left(
    \frac{e}{\omega_{n;j}}
    \right)
    E^{\alpha_{j}}(\omega_{n;j})
    c^{\dagger}_{\bm{k}a}
    h_{\bm{k}aa'}^{\mu\alpha_{1}...\alpha_{m}}
    c_{\bm{k}a'}
    \right\}.
\end{align}
Thus,
\begin{align}
    \langle\widehat{J}^{\mu}\rangle (\tau)
    =&
    \frac{e}{Z}
    \left[
    \sum_{aa'}
    \int\frac{d^{d}\bm{k}}{(2\pi)^{d}}
    \left\{
    \frac{-\delta}{\delta\gamma_{\bm{k}a}(\tau)}
    h_{\bm{k}aa'}^{\mu}
    \frac{\delta}{\delta\overline{\alpha}_{\bm{k}a'}(\tau)}
    \right.
    \right.\nonumber
    \\
    &\;\;\;\;\;\;\;\;\;\;\;\;\;\;\;\;\;\;\;\;+
    \left.
    \left.
    \sum_{m=1}^{\infty}
    \frac{1}{m!}
    \prod_{j=1}^{m}
    i\int d\omega_{n;j}
    e^{-i\omega_{n;j} \tau}
    \left(
    \frac{e}{\omega_{n;j}}
    \right)
    E^{\alpha_{j}}(\omega_{n;j})
    \frac{-\delta}{\delta\gamma_{\bm{k}a}(\tau)}
    h_{\bm{k}aa'}^{\mu\alpha_{1}...\alpha_{m}}
    \frac{\delta}{\delta\overline{\alpha}_{\bm{k}a'}(\tau)}
    \right\}\nonumber
    \right]\\
    &\times
    \mathrm{exp}
    \left[-
    \int_{0}^{\beta} d\tau\, 
    \sum_{aa'}
    \int\frac{d^{d}\bm{k}}{(2\pi)^{d}}
    \sum_{m=1}^{\infty}
    \frac{1}{m!}
    \prod_{j=1}^{m}
    i\int d\omega_{n;j}
    e^{-i\omega_{n;j} \tau}
    \right.\nonumber
    \\
    &\;\;\;\;\;\;\;\;\;\;\;\;\;\;\;\;\;\;\;\;\times
    \left.
    \left(
    \frac{e}{\omega_{n;j}}
    \right)
    E^{\alpha_{k}}(\omega_{n;j})
    \frac{-\delta}{\delta\gamma_{\bm{k}a}(\tau)}
    h_{\bm{k}aa'}^{\alpha_{1}...\alpha_{m}}
    \frac{\delta}{\delta\overline{\alpha}_{\bm{k}a'}(\tau)}
    \right]
    \left.
    Z_{0,\overline{\alpha},\gamma}
    \right|_{\overline{\alpha}=\gamma=0}.
\end{align}
The first $[\cdots]$ part can be expanded as below:
\begin{align}
    &
    \sum_{aa'}
    \int\frac{d^{d}\bm{k}}{(2\pi)^{d}}
    \left\{
    \frac{-\delta}{\delta\gamma_{\bm{k}a}(\tau)}
    h_{\bm{k}aa'}^{\mu}
    \frac{\delta}{\delta\overline{\alpha}_{\bm{k}a'}(\tau)}
    \right.\nonumber
    \\
    &\;\;\;\;\;\;\;\;\;\;+
    \left.
    \sum_{m=1}^{\infty}
    \frac{1}{m!}
    \prod_{j=1}^{m}
    i\int d\omega_{n;j}
    e^{-i\omega_{n;j} \tau}
    \left(
    \frac{e}{\omega_{n;j}}
    \right)
    E^{\alpha_{j}}(\omega_{n;j})
    \frac{-\delta}{\delta\gamma_{\bm{k}a}(\tau)}
    h_{\bm{k}aa'}^{\mu\alpha_{1}...\alpha_{m}}
    \frac{\delta}{\delta\overline{\alpha}_{\bm{k}a'}(\tau)}
    \right\}\nonumber
    \\
    =&
    \sum_{aa'}
    \int\frac{d^{d}\bm{k}}{(2\pi)^{d}}
    \left\{
    \frac{-\delta}{\delta\gamma_{\bm{k}a}(\tau)}
    h_{\bm{k}aa'}^{\mu}
    \frac{\delta}{\delta\overline{\alpha}_{\bm{k}a'}(\tau)}
    \right.\nonumber
    \\
    &\;\;\;\;\;\;\;\;\;\;+
    i\int d\omega_{n;1}
    e^{-i\omega_{n;1} \tau}
    \left(
    \frac{e}{\omega_{n;1}}
    \right)
    E^{\alpha_{1}}(\omega_{n;1})
    \frac{-\delta}{\delta\gamma_{\bm{k}a}(\tau)}
    h_{\bm{k}aa'}^{\mu\alpha_{1}}
    \frac{\delta}{\delta\overline{\alpha}_{\bm{k}a'}(\tau)}\nonumber
    \\
    &\;\;\;\;\;\;\;\;\;\;+
    \left.
    \frac{1}{2!}
    i^{2}\int d\omega_{n;1}d\omega_{n;2}
    e^{-i(\omega_{n;1}+\omega_{n;2}) \tau}
    \left(
    \frac{e^{2}}{\omega_{n;1}\omega_{n;2}}
    \right)
    E^{\alpha_{1}}(\omega_{n;1})E^{\alpha_{2}}(\omega_{n;2})
    \frac{-\delta}{\delta\gamma_{\bm{k}a}(\tau)}
    h_{\bm{k}aa'}^{\mu\alpha_{1}\alpha_{2}}
    \frac{\delta}{\delta\overline{\alpha}_{\bm{k}a'}(\tau)}
    \right\}\nonumber
    \\
    &\;\;\;\;\;\;\;\;\;\;+
    \mathcal{O}(E^{3}).
    \label{eq:J_expand_former}
\end{align}
The exponential part can be expanded as below:
\begin{align}
    &\mathrm{exp}
    \left[-
    \int_{0}^{\beta} d\tau\, 
    \sum_{aa'}
    \int\frac{d^{d}\bm{k}}{(2\pi)^{d}}
    \sum_{m=1}^{\infty}
    \frac{1}{m!}
    \prod_{j=1}^{m}
    i\int d\omega_{n;j}
    e^{-i\omega_{n;j} \tau}
    \left(
    \frac{e}{\omega_{n;j}}
    \right)
    E^{\alpha_{k}}(\omega_{n;j})
    \frac{-\delta}{\delta\gamma_{\bm{k}a}(\tau)}
    h_{\bm{k}aa'}^{\alpha_{1}...\alpha_{m}}
    \frac{\delta}{\delta\overline{\alpha}_{\bm{k}a'}(\tau)}
    \right]\nonumber
    \\
    =&
    \mathrm{exp}
    \left[
    -
    \int_{0}^{\beta} d\tau\, 
    \sum_{aa'}
    \int\frac{d^{d}\bm{k}}{(2\pi)^{d}}
    \left\{
    i\int d\omega_{n;1}
    e^{-i\omega_{n;1} \tau}
    \left(
    \frac{e}{\omega_{n;1}}
    \right)
    E^{\alpha_{1}}(\omega_{n;1})
    \frac{-\delta}{\delta\gamma_{\bm{k}a}(\tau)}
    h_{\bm{k}aa'}^{\alpha_{1}}
    \frac{\delta}{\delta\overline{\alpha}_{\bm{k}a'}(\tau)}
    \right.
    \right.\nonumber
    \\
    &\;+
    \left.
    \left.
    \frac{1}{2!}
    i^{2}\int d\omega_{n;1}d\omega_{n;2}
    e^{-i(\omega_{n;1}+\omega_{n;2}) \tau}
    \left(
    \frac{e^{2}}{\omega_{n;1}\omega_{n;2}}
    \right)
    E^{\alpha_{k}}(\omega_{n;1})
    E^{\alpha_{k}}(\omega_{n;2})
    \frac{-\delta}{\delta\gamma_{\bm{k}a}(\tau)}
    h_{\bm{k}aa'}^{\alpha_{1}\alpha_{2}}
    \frac{\delta}{\delta\overline{\alpha}_{\bm{k}a'}(\tau)}
    +\cdots
    \right\}
    \right]\nonumber
    \\
    =&\;
    1
    -
    \int_{0}^{\beta} d\tau\, 
    \sum_{aa'}
    \int\frac{d^{d}\bm{k}}{(2\pi)^{d}}
    i\int d\omega_{n;1}
    e^{-i\omega_{n;1} \tau}
    \left(
    \frac{e}{\omega_{n;1}}
    \right)
    E^{\alpha_{1}}(\omega_{n;1})
    \frac{-\delta}{\delta\gamma_{\bm{k}a}(\tau)}
    h_{\bm{k}aa'}^{\alpha_{1}}
    \frac{\delta}{\delta\overline{\alpha}_{\bm{k}a'}(\tau)}\nonumber
    \\
    &-
    \int_{0}^{\beta} d\tau\, 
    \sum_{aa'}
    \int\frac{d^{d}\bm{k}}{(2\pi)^{d}}
    \frac{1}{2!}
    i^{2}\int d\omega_{n;1}d\omega_{n;2}
    e^{-i(\omega_{n;1}+\omega_{n;2}) \tau}\nonumber
    \\
    &\;\;\;\;\;\;\;\;\;\;\;\;\;\;\;\;\;\;\;\;\times
    \left(
    \frac{e^{2}}{\omega_{n;1}\omega_{n;2}}
    \right)
    E^{\alpha_{1}}(\omega_{n;1})
    E^{\alpha_{2}}(\omega_{n;2})
    \frac{-\delta}{\delta\gamma_{\bm{k}a}(\tau)}
    h_{\bm{k}aa'}^{\alpha_{1}\alpha_{2}}
    \frac{\delta}{\delta\overline{\alpha}_{\bm{k}a'}(\tau)}\nonumber
    \\
    &+
    \frac{1}{2!}
    \int_{0}^{\beta} d\tau d\tau '\, 
    \sum_{aa'bb'}
    \int\frac{d^{d}\bm{k}}{(2\pi)^{d}}\frac{d^{d}\bm{k}'}{(2\pi)^{d}}
    i^{2}\int d\omega_{n;1}d\omega_{n;2}
    e^{-i(\omega_{n;1} \tau+\omega_{n;2} \tau ')}\nonumber
    \\
    &\;\;\;\;\;\;\;\;\;\;\;\;\;\;\;\;\;\;\;\;\times
    \left(
    \frac{e^{2}}{\omega_{n;1}\omega_{n;2}}
    \right)
    E^{\alpha_{1}}(\omega_{n;1})E^{\alpha_{2}}(\omega_{n;2})
    \frac{-\delta}{\delta\gamma_{\bm{k}a}(\tau)}
    h_{\bm{k}aa'}^{\alpha_{1}}
    \frac{\delta}{\delta\overline{\alpha}_{\bm{k}a'}(\tau)}
    \frac{-\delta}{\delta\gamma_{\bm{k}'b}(\tau ')}
    h_{\bm{k}'bb'}^{\alpha_{2}}
    \frac{\delta}{\delta\overline{\alpha}_{\bm{k}'b'}(\tau ')}\nonumber
    \\
    &+
    \mathcal{O}(E^{3}).
    \label{eq:J_expand_latter}
\end{align}
As the diagram calculation in the quantum field theory, we can express each contribution as diagrams. Here is the summary of the representation method.
\begin{itemize}
    \item The contraction of $\frac{\delta}{\delta\overline{\alpha}_{\bm{k}a}(\tau )},\;\frac{\delta}{\delta\gamma_{\bm{k}'a'}(\tau ')}$ corresponds to the line which is called ``propagator''.
    \item Such contraction is replaced by $\frac{\delta}{\delta\overline{\alpha}_{\bm{k}a}(\tau )}\frac{\delta}{\delta\gamma_{\bm{k}'a'}(\tau ')}\rightarrow(2\pi)^{d}\delta(\bm{k}-\bm{k}')\delta_{aa'}G_{\bm{k}aa}(\tau-\tau')$
    \footnote{If $\bm{k}\neq\bm{k}'$, then there remains $\overline{\alpha}G,\;G\gamma$ terms which goes to $0$ by $\overline{\alpha},\gamma\rightarrow 0$. The factor $\delta_{aa'}$ comes from \eqref{eq:gleen_fourier_final}. We must exchange the location of functional derivatives (anti-commutation!) and realize the ordering $\delta\overline{\alpha}\,\delta\gamma$ to apply \eqref{eq:gleen_partial} directly.}.
    \item The coefficient $h^{\alpha_{1}\cdots\alpha_{m}}_{\bm{k}aa'}$ corresponds to the node which is called ``vertex''.
    \item Such vertex connects propagators which represent the contraction of its indices.
    \item The external field $E^{\alpha}$ and the output current $J^{\mu}$ correspond to the line which is called ``external line''.
    \item The external line grows from the vertex which has the same vector index.
    \item The diagram which doesn't contain $\mu$ (the vector index of output current $J^{\mu}$) is called ``bubble diagram''.
    \item Applying the diagram representation, Eq.~\eqref{eq:path_integral_current} becomes
    \begin{align}
        \langle\widehat{J}^{\mu}\rangle (\tau)
        =
        \frac{(\textrm{Bubble diagrams})\times(\textrm{Non-bubble diagrams})}{(\textrm{Bubble diagrams})}.
    \end{align}
    \item Therefore, only we have to do is to calculate the contribution of ``non-bubble diagrams''.
\end{itemize}

\subsection{The first order}
The corresponding terms in the product of \eqref{eq:J_expand_former} and \eqref{eq:J_expand_latter} are
\begin{align}
    (\mathcal{O}(E)&\textrm{ terms})\nonumber
    \\
    =&\left\{
    \sum_{aa'}
    \int\frac{d^{d}\bm{k}}{(2\pi)^{d}}
    \frac{-\delta}{\delta\gamma_{\bm{k}a}(\tau)}
    h_{\bm{k}aa'}^{\mu}
    \frac{\delta}{\delta\overline{\alpha}_{\bm{k}a'}(\tau)}
    \right\}\nonumber
    \\
    &\;\;\;\;\;\;\;\;\;\;\times
    \left\{
    -
    \int_{0}^{\beta} d\tau'\, 
    \sum_{bb'}
    \int\frac{d^{d}\bm{k'}}{(2\pi)^{d}}
    i\int d\omega_{n;1}
    e^{-i\omega_{n;1} \tau'}
    \left(
    \frac{e}{\omega_{n;1}}
    \right)
    E^{\alpha_{1}}(\omega_{n;1})
    \frac{-\delta}{\delta\gamma_{\bm{k'}b}(\tau')}
    h_{\bm{k'}bb'}^{\alpha_{1}}
    \frac{\delta}{\delta\overline{\alpha}_{\bm{k'}b'}(\tau')}
    \right\}\nonumber
    \\
    &+
    \left\{
    \sum_{aa'}
    \int\frac{d^{d}\bm{k}}{(2\pi)^{d}}
    i\int d\omega_{n;1}
    e^{-i\omega_{n;1} \tau}
    \left(
    \frac{e}{\omega_{n;1}}
    \right)
    E^{\alpha_{1}}(\omega_{n;1})
    \frac{-\delta}{\delta\gamma_{\bm{k}a}(\tau)}
    h_{\bm{k}aa'}^{\mu\alpha_{1}}
    \frac{\delta}{\delta\overline{\alpha}_{\bm{k}a'}(\tau)}
    \right\}
    \times
    1\nonumber
    \\
    =&
    \sum_{ab}
    \int\frac{d^{d}\bm{k}}{(2\pi)^{d}}
    \int_{0}^{\beta} d\tau'\, 
    i\int d\omega_{n;1}
    e^{-i\omega_{n;1} \tau'}
    \left(
    \frac{e}{\omega_{n;1}}
    \right)
    E^{\alpha_{1}}(\omega_{n;1})
    h_{\bm{k}ab}^{\mu}
    G_{\bm{k}bb}(\tau-\tau')
    h_{\bm{k}ba}^{\alpha_{1}}
    G_{\bm{k}aa}(\tau'-\tau)
    \label{eq:first_order_1}
    \\
    &+
    \sum_{a}
    \int\frac{d^{d}\bm{k}}{(2\pi)^{d}}
    i\int d\omega_{n;1}
    e^{-i\omega_{n;1} \tau}
    \left(
    \frac{e}{\omega_{n;1}}
    \right)
    E^{\alpha_{1}}(\omega_{n;1})
    h_{\bm{k}aa}^{\mu\alpha_{1}}
    G_{\bm{k}aa}(\tau-\tau).
    \label{eq:first_order_2}
\end{align}

We perform the inverse Fourier transformation $\langle\widehat{J}^{\mu}\rangle(\omega_{n})=\int_{0}^{\beta}d\tau\langle\widehat{J}^{\mu}\rangle(\tau)e^{i\omega_{n}\tau}$.
Then, the relevant part of the $\tau,\;\tau'$ integral in \eqref{eq:first_order_1} becomes 
\begin{align}
    &\int_{0}^{\beta}d\tau
    e^{i\omega_{n}\tau}
    \int_{0}^{\beta} d\tau'
    e^{-i\omega_{n;1} \tau'}
    G_{\bm{k}bb}(\tau-\tau')
    G_{\bm{k}aa}(\tau'-\tau)
    =
    \sum_{n'}
    \delta_{\omega_{n},\omega_{n;1}}
    \frac{1}{i\omega_{n;1}+i\omega_{n'}-\varepsilon_{\bm{k}b}}
    \frac{1}{i\omega_{n'}-\varepsilon_{\bm{k}a}}.
\end{align}
Similarly, the relevant part of the $\tau,\;\tau'$ integral in \eqref{eq:first_order_2} becomes
\begin{align}
    &\int_{0}^{\beta}d\tau
    e^{i\omega_{n}\tau}
    e^{-i\omega_{n;1} \tau}
    G_{\bm{k}aa}(\tau-\tau)
    =
    \sum_{n'}
    \delta_{\omega_{n},\omega_{n;1}}
    \frac{1}{i\omega_{n'}-\varepsilon_{\bm{k}a}}.
\end{align}
Therefore, the first order of the output current $\langle\widehat{J}^{\mu}\rangle^{(1)}(\omega_{n})$ is
\begin{align}
    \langle\widehat{J}^{\mu}\rangle^{(1)}(\omega_{n})
    =&\;
    e\sum_{ab}
    \int\frac{d^{d}\bm{k}}{(2\pi)^{d}}
    i\int d\omega_{n;1}
    \left(
    \frac{e}{\omega_{n;1}}
    \right)
    E^{\alpha_{1}}(\omega_{n;1})\nonumber
    \\
    &\;\;\;\;\;\;\;\;\;\;\times
    \sum_{n'}
    \delta_{\omega_{n},\omega_{n;1}}
    \left[
    h_{\bm{k}ab}^{\mu}
    h_{\bm{k}ba}^{\alpha_{1}}
    \frac{1}{i\omega_{n;1}+i\omega_{n'}-\varepsilon_{\bm{k}b}}
    \frac{1}{i\omega_{n'}-\varepsilon_{\bm{k}a}}
    +
    h_{\bm{k}aa}^{\mu\alpha_{1}}
    \frac{1}{i\omega_{n'}-\varepsilon_{\bm{k}a}}
    \right].
\end{align}

\subsection{The second order}
The corresponding terms in the product of \eqref{eq:J_expand_former} and \eqref{eq:J_expand_latter} are 
\begin{align}
    (\mathcal{O}(E^{2})&\textrm{ terms})\nonumber
    \\
    =&
    \sum_{aa'}
    \int\frac{d^{d}\bm{k}}{(2\pi)^{d}}
    \left\{
    \frac{-\delta}{\delta\gamma_{\bm{k}a}(\tau)}
    h_{\bm{k}aa'}^{\mu}
    \frac{\delta}{\delta\overline{\alpha}_{\bm{k}a'}(\tau)}
    \right\}\nonumber
    \\
    &\;\;\;\;\;\;\;\;\times
    \left\{
    \frac{1}{2!}
    \int_{0}^{\beta} d\tau' d\tau''\, 
    \sum_{bb'cc'}
    \int\frac{d^{d}\bm{k}'}{(2\pi)^{d}}\frac{d^{d}\bm{k}''}{(2\pi)^{d}}
    i^{2}\int d\omega_{n;1}d\omega_{n;2}
    e^{-i(\omega_{n;1} \tau'+\omega_{n;2} \tau'')}
    \right.\nonumber
    \\
    &\;\;\;\;\;\;\;\;\times
    \left.
    \left(
    \frac{e^{2}}{\omega_{n;1}\omega_{n;2}}
    \right)
    E^{\alpha_{1}}(\omega_{n;1})E^{\alpha_{2}}(\omega_{n;2})
    \frac{-\delta}{\delta\gamma_{\bm{k}'b}(\tau')}
    h_{\bm{k}'bb'}^{\alpha_{1}}
    \frac{\delta}{\delta\overline{\alpha}_{\bm{k}'b'}(\tau')}
    \frac{-\delta}{\delta\gamma_{\bm{k}''c}(\tau'')}
    h_{\bm{k}''cc'}^{\alpha_{2}}
    \frac{\delta}{\delta\overline{\alpha}_{\bm{k}''c'}(\tau'')}
    \right\}\nonumber
    \\
    &-
    \sum_{aa'}
    \int\frac{d^{d}\bm{k}}{(2\pi)^{d}}
    \left\{
    \frac{-\delta}{\delta\gamma_{\bm{k}a}(\tau)}
    h_{\bm{k}aa'}^{\mu}
    \frac{\delta}{\delta\overline{\alpha}_{\bm{k}a'}(\tau)}
    \right\}\nonumber
    \\
    &\;\;\;\;\;\;\;\;\times
    \left\{
    \int_{0}^{\beta} d\tau'\, 
    \sum_{bb'}
    \int\frac{d^{d}\bm{k}'}{(2\pi)^{d}}
    \frac{1}{2!}
    i^{2}\int d\omega_{n;1}d\omega_{n;2}
    e^{-i(\omega_{n;1}+\omega_{n;2}) \tau'}
    \right.\nonumber
    \\
    &\;\;\;\;\;\;\;\;\times
    \left.
    \left(
    \frac{e^{2}}{\omega_{n;1}\omega_{n;2}}
    \right)
    E^{\alpha_{1}}(\omega_{n;1})
    E^{\alpha_{2}}(\omega_{n;2})
    \frac{-\delta}{\delta\gamma_{\bm{k}'b}(\tau')}
    h_{\bm{k}'bb'}^{\alpha_{1}\alpha_{2}}
    \frac{\delta}{\delta\overline{\alpha}_{\bm{k}'b'}(\tau')}
    \right\}\nonumber
    \\
    &-
    \sum_{aa'}
    \int\frac{d^{d}\bm{k}}{(2\pi)^{d}}
    \left\{
    i\int d\omega_{n;1}
    e^{-i\omega_{n;1} \tau}
    \left(
    \frac{e}{\omega_{n;1}}
    \right)
    E^{\alpha_{1}}(\omega_{n;1})
    \frac{-\delta}{\delta\gamma_{\bm{k}a}(\tau)}
    h_{\bm{k}aa'}^{\mu\alpha_{1}}
    \frac{\delta}{\delta\overline{\alpha}_{\bm{k}a'}(\tau)}
    \right\}\nonumber
    \\
    &\;\;\;\;\;\;\;\;\times
    \left\{
    \int_{0}^{\beta} d\tau'\, 
    \sum_{bb'}
    \int\frac{d^{d}\bm{k}'}{(2\pi)^{d}}
    i\int d\omega_{n;2}
    e^{-i\omega_{n;2} \tau'}
    \left(
    \frac{e}{\omega_{n;2}}
    \right)
    E^{\alpha_{2}}(\omega_{n;2})
    \frac{-\delta}{\delta\gamma_{\bm{k}'b'}(\tau')}
    h_{\bm{k}'bb'}^{\alpha_{2}}
    \frac{\delta}{\delta\overline{\alpha}_{\bm{k}'b}(\tau')}
    \right\}\nonumber
    \\
    &+
    \sum_{aa'}
    \int\frac{d^{d}\bm{k}}{(2\pi)^{d}}
    \left\{
    \frac{1}{2!}
    i^{2}\int d\omega_{n;1}d\omega_{n;2}
    e^{-i(\omega_{n;1}+\omega_{n;2}) \tau}
    \right.\nonumber
    \\
    &\;\;\;\;\;\;\;\;\times
    \left.
    \left(
    \frac{e^{2}}{\omega_{n;1}\omega_{n;2}}
    \right)
    E^{\alpha_{1}}(\omega_{n;1})E^{\alpha_{2}}(\omega_{n;2})
    \frac{-\delta}{\delta\gamma_{\bm{k}a'}(\tau)}
    h_{\bm{k}aa'}^{\mu\alpha_{1}\alpha_{2}}
    \frac{\delta}{\delta\overline{\alpha}_{\bm{k}a}(\tau)}
    \right\}
    \times 1\nonumber
    \\
    =&
    \frac{1}{2!}
    \sum_{abc}
    \int
    \frac{d^{d}\bm{k}}{(2\pi)^{d}}
    \int_{0}^{\beta} d\tau' d\tau''\, 
    i^{2}\int d\omega_{n;1}d\omega_{n;2}
    e^{-i(\omega_{n;1} \tau'+\omega_{n;2} \tau'')}\nonumber
    \\
    &\;\;\;\;\;\;\;\;\;\;\times
    \left(
    \frac{e^{2}}{\omega_{n;1}\omega_{n;2}}
    \right)
    E^{\alpha_{1}}(\omega_{n;1})E^{\alpha_{2}}(\omega_{n;2})
    h_{\bm{k}ab}^{\mu}
    G_{\bm{k}bb}(\tau-\tau')
    h_{\bm{k}bc}^{\alpha_{1}}
    G_{\bm{k}cc}(\tau'-\tau'')
    h_{\bm{k}ca}^{\alpha_{2}}
    G_{\bm{k}aa}(\tau''-\tau)
    \label{eq:second_order_1}
    \\
    &+
    [(\omega_{n;1},\alpha_{1})\longleftrightarrow(\omega_{n;2},\alpha_{2})]\nonumber
    \\
    &+
    \frac{1}{2!}
    \sum_{ab}
    \int
    \frac{d^{d}\bm{k}}{(2\pi)^{d}}
    \int_{0}^{\beta} d\tau'\, 
    i^{2}\int d\omega_{n;1}d\omega_{n;2}
    e^{-i(\omega_{n;1}+\omega_{n;2})\tau'}\nonumber
    \\
    &\;\;\;\;\;\;\;\;\;\;\times
    \left(
    \frac{e^{2}}{\omega_{n;1}\omega_{n;2}}
    \right)
    E^{\alpha_{1}}(\omega_{n;1})E^{\alpha_{2}}(\omega_{n;2})
    h_{\bm{k}ab}^{\mu}
    G_{\bm{k}bb}(\tau-\tau')
    h_{\bm{k}ba}^{\alpha_{1}\alpha_{2}}
    G_{\bm{k}aa}(\tau'-\tau)
    \label{eq:second_order_2}
    \\
    &+
    \sum_{ab}
    \int
    \frac{d^{d}\bm{k}}{(2\pi)^{d}}
    \int_{0}^{\beta} d\tau'\, 
    i^{2}\int d\omega_{n;1}d\omega_{n;2}
    e^{-i(\omega_{n;1}\tau+\omega_{n;2}\tau')}\nonumber
    \\
    &\;\;\;\;\;\;\;\;\;\;\times
    \left(
    \frac{e^{2}}{\omega_{n;1}\omega_{n;2}}
    \right)
    E^{\alpha_{1}}(\omega_{n;1})E^{\alpha_{2}}(\omega_{n;2})
    h_{\bm{k}ab}^{\mu\alpha_{1}}
    G_{\bm{k}bb}(\tau-\tau')
    h_{\bm{k}ba}^{\alpha_{2}}
    G_{\bm{k}aa}(\tau'-\tau)
    \label{eq:second_order_3}
    \\
    &+
    \frac{1}{2!}
    \sum_{a}
    \int
    \frac{d^{d}\bm{k}}{(2\pi)^{d}}
    i^{2}\int d\omega_{n;1}d\omega_{n;2}
    e^{-i(\omega_{n;1}+\omega_{n;2})\tau}\nonumber
    \\
    &\;\;\;\;\;\;\;\;\;\;\times
    \left(
    \frac{e^{2}}{\omega_{n;1}\omega_{n;2}}
    \right)
    E^{\alpha_{1}}(\omega_{n;1})E^{\alpha_{2}}(\omega_{n;2})
    h_{\bm{k}aa}^{\mu\alpha_{1}\alpha_{2}}
    G_{\bm{k}aa}(\tau-\tau).
    \label{eq:second_order_4}
\end{align}
Performing the inverse Fourier transformation $\langle\widehat{J}^{\mu}\rangle(\omega_{n})=\int_{0}^{\beta}d\tau\langle\widehat{J}^{\mu}\rangle(\tau)e^{i\omega_{n}\tau}$, the relevant part of the $\tau,\;\tau'$ integral in \eqref{eq:second_order_1} becomes
\begin{align}
    &
    \int_{0}^{\beta}d\tau
    e^{i\omega_{n}\tau}
    \int_{0}^{\beta} d\tau' d\tau''\, 
    e^{-i(\omega_{n;1} \tau'+\omega_{n;2} \tau'')}
    G_{\bm{k}bb}(\tau-\tau')
    G_{\bm{k}cc}(\tau'-\tau'')
    G_{\bm{k}aa}(\tau''-\tau)\nonumber
    \\
    =&
    \sum_{n'}
    \delta_{\omega_{n},\omega_{n;2}+\omega_{n;1}}
    \frac{1}{i(\omega_{n;1}+\omega_{n;2}+\omega_{n'})-\varepsilon_{\bm{k}b}}
    \frac{1}{i(\omega_{n;2}+\omega_{n'})-\varepsilon_{\bm{k}c}}
    \frac{1}{i\omega_{n'}-\varepsilon_{\bm{k}a}}.
\end{align}
Similarly, for \eqref{eq:second_order_2},
\begin{align}
    &
    \int_{0}^{\beta}d\tau
    e^{i\omega_{n}\tau}
    \int_{0}^{\beta} d\tau'\, 
    e^{-i(\omega_{n;1}+\omega_{n;2})\tau'}
    G_{\bm{k}bb}(\tau-\tau')
    G_{\bm{k}aa}(\tau'-\tau)\nonumber
    \\
    =&
    \sum_{n'}
    \delta_{\omega_{n},\omega_{n;1}+\omega_{n;2}}
    \frac{1}{i(\omega_{n;1}+\omega_{n;2}+\omega_{n'})-\varepsilon_{\bm{k}b}}
    \frac{1}{i\omega_{n'}-\varepsilon_{\bm{k}a}}.
\end{align}
For \eqref{eq:second_order_3},
\begin{align}
    &
    \int_{0}^{\beta}d\tau
    e^{i\omega_{n}\tau}
    \int_{0}^{\beta} d\tau'\, 
    e^{-i(\omega_{n;1}\tau+\omega_{n;2}\tau')}
    G_{\bm{k}bb}(\tau-\tau')
    G_{\bm{k}aa}(\tau'-\tau)\nonumber
    \\
    =&
    \sum_{n'}
    \delta_{\omega_{n},\omega_{n;1}+\omega_{n;2}}
    \frac{1}{i(\omega_{n;2}+\omega_{n'})-\varepsilon_{\bm{k}b}}
    \frac{1}{i\omega_{n'}-\varepsilon_{\bm{k}a}}.
\end{align}
For \eqref{eq:second_order_4},
\begin{align}
    &
    \int_{0}^{\beta}d\tau
    e^{i\omega_{n}\tau}
    e^{-i(\omega_{n;1}+\omega_{n;2})\tau}
    G_{\bm{k}aa}(\tau-\tau)
    =
    \sum_{n'}
    \delta_{\omega_{n},\omega_{n;1}+\omega_{n;2}}
    \frac{1}{i\omega_{n'}-\varepsilon_{\bm{k}a}}.
\end{align}
Therefore, the second order of the output current $\langle\widehat{J}^{\mu}\rangle^{(2)}(\omega_{n})$ is
\begin{align}
    \langle\widehat{J}^{\mu}\rangle^{(2)}(\omega_{n})
    =&\;
    e
    \frac{1}{2!}
    \sum_{abc}
    \int
    \frac{d^{d}\bm{k}}{(2\pi)^{d}}
    i^{2}\int d\omega_{n;1}d\omega_{n;2}
    \left(
    \frac{e^{2}}{\omega_{n;1}\omega_{n;2}}
    \right)
    E^{\alpha_{1}}(\omega_{n;1})E^{\alpha_{2}}(\omega_{n;2})
    \sum_{n'}
    \delta_{\omega_{n},\omega_{n;1}+\omega_{n;2}}\nonumber
    \\
    &\;\;\;\;\;\;\;\;\;\;\;\;\;\;\;\times
    h_{\bm{k}ab}^{\mu}
    h_{\bm{k}bc}^{\alpha_{1}}
    h_{\bm{k}ca}^{\alpha_{2}}
    \frac{1}{i(\omega_{n;1}+\omega_{n;2}+\omega_{n'})-\varepsilon_{\bm{k}b}}
    \frac{1}{i(\omega_{n;2}+\omega_{n'})-\varepsilon_{\bm{k}c}}
    \frac{1}{i\omega_{n'}-\varepsilon_{\bm{k}a}}\nonumber
    \\
    &+
    [(\omega_{n;1},\alpha_{1})\longleftrightarrow(\omega_{n;2},\alpha_{2})]\nonumber
    \\
    &+
    e
    \frac{1}{2!}
    \sum_{ab}
    \int
    \frac{d^{d}\bm{k}}{(2\pi)^{d}}
    i^{2}\int d\omega_{n;1}d\omega_{n;2}
    \left(
    \frac{e^{2}}{\omega_{n;1}\omega_{n;2}}
    \right)
    \sum_{n'}
    \delta_{\omega_{n},\omega_{n;1}+\omega_{n;2}}
    E^{\alpha_{1}}(\omega_{n;1})E^{\alpha_{2}}(\omega_{n;2})\nonumber
    \\
    &\;\;\;\;\;\;\;\;\;\;\;\;\;\;\;\times
    h_{\bm{k}ab}^{\mu}
    h_{\bm{k}ba}^{\alpha_{1}\alpha_{2}}
    \frac{1}{i(\omega_{n;1}+\omega_{n;2}+\omega_{n'})-\varepsilon_{\bm{k}b}}
    \frac{1}{i\omega_{n'}-\varepsilon_{\bm{k}a}}\nonumber
    \\
    &+
    e
    \sum_{ab}
    \int
    \frac{d^{d}\bm{k}}{(2\pi)^{d}}
    i^{2}\int d\omega_{n;1}d\omega_{n;2}
    \left(
    \frac{e^{2}}{\omega_{n;1}\omega_{n;2}}
    \right)
    E^{\alpha_{1}}(\omega_{n;1})E^{\alpha_{2}}(\omega_{n;2})
    \sum_{n'}
    \delta_{\omega_{n},\omega_{n;1}+\omega_{n;2}}\nonumber
    \\
    &\;\;\;\;\;\;\;\;\;\;\;\;\;\;\;\times
    h_{\bm{k}ab}^{\mu\alpha_{1}}
    h_{\bm{k}ba}^{\alpha_{2}}
    \frac{1}{i(\omega_{n;2}+\omega_{n'})-\varepsilon_{\bm{k}b}}
    \frac{1}{i\omega_{n'}-\varepsilon_{\bm{k}a}}\nonumber
    \\
    &+
    e
    \frac{1}{2!}
    \sum_{a}
    \int
    \frac{d^{d}\bm{k}}{(2\pi)^{d}}
    i^{2}\int d\omega_{n;1}d\omega_{n;2}
    \left(
    \frac{e^{2}}{\omega_{n;1}\omega_{n;2}}
    \right)
    E^{\alpha_{1}}(\omega_{n;1})E^{\alpha_{2}}(\omega_{n;2})
    \sum_{n'}
    \delta_{\omega_{n},\omega_{n;1}+\omega_{n;2}}\nonumber
    \\
    &\;\;\;\;\;\;\;\;\;\;\;\;\;\;\;\times
    h_{\bm{k}aa}^{\mu\alpha_{1}\alpha_{2}}
    \frac{1}{i\omega_{n'}-\varepsilon_{\bm{k}a}}.
\end{align}
Finally, we go back to the $\{t,\omega\}$ representation
\begin{align}
    \langle\widehat{J}^{\mu}\rangle^{(2)}(\omega)
    =&\;
    e
    \frac{1}{2!}
    \sum_{abc}
    \int
    \frac{d^{d}\bm{k}}{(2\pi)^{d}}
    \int d\omega_{1}d\omega_{2}
    \left(
    \frac{i^{2}e^{2}}{\omega_{1}\omega_{2}}
    \right)
    E^{\alpha_{1}}(\omega_{1})E^{\alpha_{2}}(\omega_{2})
    \int d\omega'
    \delta(\omega-\omega_{1}-\omega_{2})\nonumber
    \\
    &\;\;\;\;\;\;\;\;\;\;\;\;\;\;\;\times
    h_{\bm{k}ab}^{\mu}
    h_{\bm{k}bc}^{\alpha_{1}}
    h_{\bm{k}ca}^{\alpha_{2}}
    \frac{1}{(\omega_{1}+\omega_{2}+\omega')-\varepsilon_{\bm{k}b}}
    \frac{1}{(\omega_{2}+\omega')-\varepsilon_{\bm{k}c}}
    \frac{1}{\omega'-\varepsilon_{\bm{k}a}}\nonumber
    \\
    &+
    e
    \frac{1}{2}\frac{1}{2!}
    \sum_{ab}
    \int
    \frac{d^{d}\bm{k}}{(2\pi)^{d}}
    \int d\omega_{1}d\omega_{2}
    \left(
    \frac{i^{2}e^{2}}{\omega_{1}\omega_{2}}
    \right)
    \int d\omega'
    \delta(\omega-\omega_{1}-\omega_{2})
    E^{\alpha_{1}}(\omega_{1})E^{\alpha_{2}}(\omega_{2})\nonumber
    \\
    &\;\;\;\;\;\;\;\;\;\;\;\;\;\;\;\times
    h_{\bm{k}ab}^{\mu}
    h_{\bm{k}ba}^{\alpha_{1}\alpha_{2}}
    \frac{1}{(\omega_{1}+\omega_{2}+\omega')-\varepsilon_{\bm{k}b}}
    \frac{1}{\omega'-\varepsilon_{\bm{k}a}}\nonumber
    \\
    &+
    e
    \frac{1}{2}
    \sum_{ab}
    \int
    \frac{d^{d}\bm{k}}{(2\pi)^{d}}
    \int d\omega_{1}d\omega_{2}
    \left(
    \frac{i^{2}e^{2}}{\omega_{1}\omega_{2}}
    \right)
    E^{\alpha_{1}}(\omega_{1})E^{\alpha_{2}}(\omega_{2})
    \int d\omega'
    \delta(\omega-\omega_{1}-\omega_{2})\nonumber
    \\
    &\;\;\;\;\;\;\;\;\;\;\;\;\;\;\;\times
    h_{\bm{k}ab}^{\mu\alpha_{1}}
    h_{\bm{k}ba}^{\alpha_{2}}
    \frac{1}{(\omega_{2}+\omega')-\varepsilon_{\bm{k}b}}
    \frac{1}{\omega'-\varepsilon_{\bm{k}a}}\nonumber
    \\
    &+
    e
    \frac{1}{2}\frac{1}{2!}
    \sum_{a}
    \int
    \frac{d^{d}\bm{k}}{(2\pi)^{d}}
    \int d\omega_{1}d\omega_{2}
    \left(
    \frac{i^{2}e^{2}}{\omega_{1}\omega_{2}}
    \right)
    E^{\alpha_{1}}(\omega_{1})E^{\alpha_{2}}(\omega_{2})
    \int d\omega'
    \delta(\omega-\omega_{1}-\omega_{2})\nonumber
    \\
    &\;\;\;\;\;\;\;\;\;\;\;\;\;\;\;\times
    h_{\bm{k}aa}^{\mu\alpha_{1}\alpha_{2}}
    \frac{1}{\omega'-\varepsilon_{\bm{k}a}}\nonumber
    \\
    &+
    [(\omega_{1},\alpha_{1})\longleftrightarrow(\omega_{2},\alpha_{2})],
    \label{eq:general_second_response}
\end{align}
where we symmetrized all of the terms. That is the reason that the factor ``$1/2$'' is added on the latter three terms and the position 
 of ``$[(\omega_{1},\alpha_{1})\longleftrightarrow(\omega_{2},\alpha_{2})]$'' is changed to the bottom.
Here is the integration formula:
\begin{align}
    I_{1}
    &=
    \int d\omega'
    \frac{1}{\omega'-\varepsilon_{\bm{k}a}}
    =f_{a}
    \\
    I_{2}(\omega_{1})
    &=
    \int d\omega'
    \frac{1}{\omega'-\varepsilon_{\bm{k}a}}
    \frac{1}{\omega'+\omega_{1}-\varepsilon_{\bm{k}b}}
    =
    \frac{f_{ab}}{\omega_{1}-\varepsilon_{\bm{k}ba}}
    \\
    I_{3}(\omega_{1},\omega_{2})
    &=
    \int d\omega'
    \frac{1}{\omega'-\varepsilon_{\bm{k}a}}
    \frac{1}{\omega'+\omega_{1}-\varepsilon_{\bm{k}b}}
    \frac{1}{\omega'+\omega_{1}+\omega_{2}-\varepsilon_{\bm{k}c}}
    =
    \frac
    {(\omega_{2}-\varepsilon_{\bm{k}cb})f_{ab}+(\omega_{1}-\varepsilon_{\bm{k}ba})f_{cb}}
    {(\omega_{1}-\varepsilon_{\bm{k}ba})(\omega_{2}-\varepsilon_{\bm{k}cb})(\omega-\varepsilon_{\bm{k}ca})},
\end{align}
where $f_{a}=f(\varepsilon_{\bm{k}a})=\frac{1}{\exp(\beta\varepsilon_{\bm{k}a})+1}$ is the Fermi factor in the finite inverse temperature $\beta$ and $f_{ab}=f_{a}-f_{b},\;\;\varepsilon_{\bm{k}ab}=\varepsilon_{\bm{k}a}-\varepsilon_{\bm{k}b}$ are differences of Fermi factors and energies, and $\omega=\omega_{1}+\omega_{2}$. Using them, the current becomes,
\begin{align}
    \langle&\widehat{J}^{\mu}\rangle^{(2)}(\omega)
    =\;
    \frac{1}{2}
    \sum_{abc}
    \int
    \frac{d^{d}\bm{k}}{(2\pi)^{d}}
    \int d\omega_{1}d\omega_{2}
    \left(
    \frac{-e^{3}}{\omega_{1}\omega_{2}}
    \right)
    E^{\alpha_{1}}(\omega_{1})E^{\alpha_{2}}(\omega_{2})
    \delta(\omega-\omega_{1}-\omega_{2})
    \nonumber\\
    &\times\left[
    h_{\bm{k}ac}^{\mu}
    h_{\bm{k}cb}^{\alpha_{1}}
    h_{\bm{k}ba}^{\alpha_{2}}
    \frac
    {(\omega_{1}-\varepsilon_{\bm{k}cb})f_{ab}+(\omega_{2}-\varepsilon_{\bm{k}ba})f_{cb}}
    {(\omega_{2}-\varepsilon_{\bm{k}ba})(\omega_{1}-\varepsilon_{\bm{k}cb})(\omega-\varepsilon_{\bm{k}ca})}
    \right.
    +
    \frac{1}{2}
    h_{\bm{k}ab}^{\mu}
    h_{\bm{k}ba}^{\alpha_{1}\alpha_{2}}
    \frac{f_{ab}}{\omega_{1}+\omega_{2}-\varepsilon_{\bm{k}ba}}
    \nonumber\\
    &\;\;\;\;\;\;
    +
    h_{\bm{k}ab}^{\mu\alpha_{1}}
    h_{\bm{k}ba}^{\alpha_{2}}
    \frac{f_{ab}}{\omega_{2}-\varepsilon_{\bm{k}ba}}
    +
    \left.
    \frac{1}{2}
    h_{\bm{k}aa}^{\mu\alpha_{1}\alpha_{2}}
    f_{a}
    +
    [(\omega_{1},\alpha_{1})\longleftrightarrow(\omega_{2},\alpha_{2})]
    \right].
    \label{eq:general_second_response2}
\end{align}
Hereafter, we reintroduce $\gamma$ by replacing $\omega_{1,2}\rightarrow\omega_{1,2}+i\gamma$.

To get $h_{\bm{k}aa'}^{\alpha_{1}...\alpha_{n}}$, we need to calculate the $k$-dependent form of the covariant derivative $\widehat{h}^{\alpha_{1}...\alpha_{n}}(\bm{k})=\widehat{\mathcal{D}}^{\alpha_{1}}\cdots\widehat{\mathcal{D}}^{\alpha_{n}}[\widehat{H}_{0}](\bm{k})$. Operating $\widehat{\bm{\mathcal{D}}}$ to $\widehat{\mathcal{O}}$ once,
\begin{align}
    \widehat{\bm{\mathcal{D}}}
    [\widehat{\mathcal{O}}](\bm{k})
    =&
    e^{-i\bm{k}\cdot\bm{r}}
    [-i\,\widehat{\bm{r}},\;\widehat{\mathcal{O}}]
    e^{i\bm{k}\cdot\bm{r}}\nonumber
    \\
    =&
    e^{-i\bm{k}\cdot\bm{r}}
    \widehat{\mathcal{O}}\,i\,\widehat{\bm{r}}
    e^{i\bm{k}\cdot\bm{r}}
    -
    e^{-i\bm{k}\cdot\bm{r}}
    i\,\widehat{\bm{r}}\widehat{\mathcal{O}}
    e^{i\bm{k}\cdot\bm{r}}\nonumber
    \\
    =&
    e^{-i\bm{k}\cdot\bm{r}}
    \widehat{\mathcal{O}}
    \left(
    \frac{\partial}{\partial\bm{k}}
    e^{i\bm{k}\cdot\bm{r}}
    \right)
    +
    \left(
    \frac{\partial}{\partial\bm{k}}
    e^{-i\bm{k}\cdot\bm{r}}
    \right)
    \widehat{\mathcal{O}}
    e^{i\bm{k}\cdot\bm{r}}\nonumber
    \\
    =&
    \frac{\partial}{\partial\bm{k}}
    \widehat{\mathcal{O}}(\bm{k}).
    \label{eq:covariant_k_once}
\end{align}
Operating $\widehat{\bm{\mathcal{D}}}$ to $\widehat{\mathcal{O}}$ twice,
\begin{align}
    \widehat{\bm{\mathcal{D}}}
    [\widehat{\bm{\mathcal{D}}}
    [\widehat{\mathcal{O}}]](\bm{k})
    =&
    \frac{\partial}{\partial\bm{k}}
    \left(
    \widehat{\bm{\mathcal{D}}}
    [\widehat{\mathcal{O}}](\bm{k})
    \right)
    =
    \frac{\partial}{\partial\bm{k}}
    \frac{\partial}{\partial\bm{k}}
    \widehat{\mathcal{O}}(\bm{k}).
\end{align}
Therefore, in $k$-dependent form, we can treat $\widehat{\bm{\mathcal{D}}}$ as a $k$-derivative.
\begin{align}
    \widehat{h}^{\alpha_{1}...\alpha_{n}}(\bm{k})
    =
    \partial_{k^{\alpha_{1}}}
    \cdots
    \partial_{k^{\alpha_{n}}}
    \widehat{H}_{0}(\bm{k}).
\end{align}
Its matrix element is,
\begin{align}
    h_{\bm{k}aa'}^{\alpha_{1}...\alpha_{n}}
    =&
    \langle u_{\bm{k}a'}|
    \partial_{k^{\alpha_{1}}}
    \cdots
    \partial_{k^{\alpha_{n}}}
    \widehat{H}_{0}(\bm{k})
    |u_{\bm{k}a}\rangle_{V}.
\end{align}
To describe it in terms of the Berry connection $\mathcal{A}^{\mu}_{ab}(\bm{k})$, we use the following relation:
\begin{align}
    \langle\psi_{\bm{k}a}|
    \mathcal{D}^{\mu}(\bm{k})
    |\psi_{\bm{k}b}\rangle
    =
    \langle\psi_{\bm{k}a}|
    \left(\delta_{ab}\partial_{k^{\mu}}
    -
    i\mathcal{A}^{\mu}(\bm{k})\right)
    |\psi_{\bm{k}b}\rangle.
\end{align}
Then, for $h_{\bm{k} ab}^\mu$,
\begin{align}
    h_{\bm{k} ab}^\mu
    =&
    \langle\psi_{\bm{k}a}|
    \left[\mathcal{D}^{\mu}(\bm{k}),\,\widehat{H}_{0}\right]
    |\psi_{\bm{k}b}\rangle\nonumber
    \\
    =&
    \langle\psi_{\bm{k}a}|
    \left(\delta_{ab}\partial_{k^{\mu}}
    -
    i\mathcal{A}^{\mu}(\bm{k})\right)
    |\psi_{\bm{k}b}\rangle\varepsilon_{\bm{k}b}
    -
    \varepsilon_{\bm{k}a}\langle\psi_{\bm{k}a}|
    \left(\delta_{ab}\partial_{k^{\mu}}
    -
    i\mathcal{A}^{\mu}(\bm{k})\right)
    |\psi_{\bm{k}b}\rangle\nonumber
    \\
    =&\,
    \delta_{ab}\partial_{k^{\mu}}\varepsilon_{\bm{k}a}
    +
    i\varepsilon_{\bm{k}ab}\mathcal{A}^\mu_{ab}(\bm{k}).
    \label{eq:h1berry}
\end{align}
For $h^{\mu \alpha}_{\bm{k}ab}$,
\begin{align}
    h^{\mu \alpha}_{\bm{k}ab}
    =&
    \langle\psi_{\bm{k}a}|
    \left[\mathcal{D}^{\mu}(\bm{k}),\,\widehat{h}^{\alpha}\right]
    |\psi_{\bm{k}b}\rangle\nonumber
    \\
    =&
    \langle\psi_{\bm{k}a}|
    \mathcal{D}^{\mu}(\bm{k})
    |\psi_{\bm{k}c}\rangle
    \langle\psi_{\bm{k}c}|
    \widehat{h}^{\alpha}
    |\psi_{\bm{k}b}\rangle
    -
    \langle\psi_{\bm{k}a}|
    \widehat{h}^{\alpha}
    |\psi_{\bm{k}c}\rangle
    \langle\psi_{\bm{k}c}|
    \mathcal{D}^{\mu}(\bm{k})
    |\psi_{\bm{k}b}\rangle\nonumber
    \\
    =&
    \langle\psi_{\bm{k}a}|
    \left(\delta_{ac}\partial_{k^{\mu}}
    -
    i\mathcal{A}^{\mu}(\bm{k})\right)
    |\psi_{\bm{k}c}\rangle
    h_{\bm{k}cb}^{\alpha}
    -
    h_{\bm{k}ac}^{\alpha}
    \langle\psi_{\bm{k}c}|
    \left(\delta_{cb}\partial_{k^{\mu}}
    -
    i\mathcal{A}^{\mu}(\bm{k})\right)
    |\psi_{\bm{k}b}\rangle\nonumber
    \\
    =&\,
    \partial_{k^{\mu}}h_{\bm{k}ab}^{\alpha}
    -
    i\mathcal{A}^{\mu}_{ac}(\bm{k})h_{\bm{k}cb}^{\alpha}
    +
    ih_{\bm{k}ac}^{\alpha}\mathcal{A}^{\mu}_{cb}(\bm{k}).
    \label{eq:h2berry}
\end{align}
\subsection{Shift current}
Shift current is the $\gamma$ independent part of the second-order response. It is obtained from the contribution satisfying $a\neq c$. We assume that $\gamma,\,\omega$ is sufficiently small compared to the energy scale of the input field. Then the resonant contribution
($\omega_{1},\,\omega_{2}\sim\varepsilon_{\bm{k}ab}$) is
\begin{align}
    \langle\widehat{J}&^{\mu}_{\mathrm{shift}}\rangle^{(2)}(\omega)\nonumber\\
    =&\;
    \frac{1}{2}
    \sum_{abcd(a\neq c)}
    \int
    \frac{d^{d}\bm{k}}{(2\pi)^{d}}
    \int d\omega_{1}d\omega_{2}
    \left(
    \frac{-e^{3}}{\omega_{1}\omega_{2}}
    \right)
    E^{\alpha_{1}}(\omega_{1})E^{\alpha_{2}}(\omega_{2})
    \delta(\omega-\omega_{1}-\omega_{2})
    \nonumber\\
    &\times\left[
    h_{\bm{k}ac}^{\mu}
    h_{\bm{k}cb}^{\alpha_{1}}
    h_{\bm{k}ba}^{\alpha_{2}}
    \frac
    {f_{ab}}
    {(\omega_{2}-\varepsilon_{\bm{k}ba}+i\gamma)(-\varepsilon_{\bm{k}ca})}
    +
    h_{\bm{k}ac}^{\mu}
    h_{\bm{k}cb}^{\alpha_{1}}
    h_{\bm{k}ba}^{\alpha_{2}}
    \frac
    {f_{cb}}
    {(\omega_{1}-\varepsilon_{\bm{k}cb}+i\gamma)(-\varepsilon_{\bm{k}ca})}
    \right.\nonumber\\
    &\;\;\;\;\;\;
    \left.
    +
    h_{\bm{k}ab}^{\mu\alpha_{1}}
    h_{\bm{k}ba}^{\alpha_{2}}
    \frac{f_{ab}}{\omega_{2}-\varepsilon_{\bm{k}ba}+i\gamma}
    +
    [(\omega_{1},\alpha_{1})\longleftrightarrow(\omega_{2},\alpha_{2})]
    \right]\nonumber
    \\
    =&\;
    \frac{1}{2}
    \sum_{abcd(a\neq c)}
    \int
    \frac{d^{d}\bm{k}}{(2\pi)^{d}}
    \int d\omega_{1}d\omega_{2}
    \left(
    \frac{-e^{3}}{\omega_{1}\omega_{2}}
    \right)
    E^{\alpha_{1}}(\omega_{1})E^{\alpha_{2}}(\omega_{2})
    \delta(\omega-\omega_{1}-\omega_{2})
    \nonumber\\
    &\times\left[
    i
    \mathcal{A}^\mu_{ac}(\bm{k})
    h_{\bm{k}cb}^{\alpha_{1}}
    h_{\bm{k}ba}^{\alpha_{2}}
    \frac
    {f_{ab}}
    {\omega_{2}-\varepsilon_{\bm{k}ba}+i\gamma}
    +
    i\mathcal{A}^\mu_{ac}(\bm{k})
    h_{\bm{k}cb}^{\alpha_{1}}
    h_{\bm{k}ba}^{\alpha_{2}}
    \frac
    {f_{cb}}
    {\omega_{1}-\varepsilon_{\bm{k}cb}+i\gamma}
    +
    \partial_{k^{\mu}}h_{\bm{k}ab}^{\alpha_{1}}
    h_{\bm{k}ba}^{\alpha_{2}}
    \frac{f_{ab}}{\omega_{2}-\varepsilon_{\bm{k}ba}+i\gamma}
    \right.\nonumber\\
    &\;\;\;\;\;\;
    \left.
    -
    i
    \mathcal{A}^{\mu}_{ad}(\bm{k})h_{\bm{k}db}^{\alpha_{1}}
    h_{\bm{k}ba}^{\alpha_{2}}
    \frac{f_{ab}}{\omega_{2}-\varepsilon_{\bm{k}ba}+i\gamma}
    +
    ih_{\bm{k}ad}^{\alpha_{1}}\mathcal{A}^{\mu}_{db}(\bm{k})
    h_{\bm{k}ba}^{\alpha_{2}}
    \frac{f_{ab}}{\omega_{2}-\varepsilon_{\bm{k}ba}+i\gamma}
    +
    [(\omega_{1},\alpha_{1})\longleftrightarrow(\omega_{2},\alpha_{2})]
    \right],
\end{align}
where we have used Eqs.~\eqref{eq:h1berry} and \eqref{eq:h2berry}. The summation over $d$ is decomposed into $d=a$ part and $d\neq a$ part. Then, for the $d\neq a$ part, dummy variables $d$ and $c$ play the same role in the expression. Therefore they are merged into a single variable $c$ to continue the calculation:
\begin{align}
    \langle\widehat{J}&^{\mu}_{\mathrm{shift}}\rangle^{(2)}(\omega)\nonumber\\
    =&\;
    \frac{1}{2}
    \sum_{abc(a\neq c)}
    \int
    \frac{d^{d}\bm{k}}{(2\pi)^{d}}
    \int d\omega_{1}d\omega_{2}
    \left(
    \frac{-e^{3}}{\omega_{1}\omega_{2}}
    \right)
    E^{\alpha_{1}}(\omega_{1})E^{\alpha_{2}}(\omega_{2})
    \delta(\omega-\omega_{1}-\omega_{2})
    \nonumber\\
    &\times\left[
    i
    \mathcal{A}^\mu_{ac}(\bm{k})
    h_{\bm{k}cb}^{\alpha_{1}}
    h_{\bm{k}ba}^{\alpha_{2}}
    \frac
    {f_{ab}}
    {\omega_{2}-\varepsilon_{\bm{k}ba}+i\gamma}
    +
    i\mathcal{A}^\mu_{ac}(\bm{k})
    h_{\bm{k}cb}^{\alpha_{1}}
    h_{\bm{k}ba}^{\alpha_{2}}
    \frac
    {f_{cb}}
    {\omega_{1}-\varepsilon_{\bm{k}cb}+i\gamma}
    \right.\nonumber\\
    &\;\;\;\;\;\;
    +
    \partial_{k^{\mu}}h_{\bm{k}ab}^{\alpha_{1}}
    h_{\bm{k}ba}^{\alpha_{2}}
    \frac{f_{ab}}{\omega_{2}-\varepsilon_{\bm{k}ba}+i\gamma}
    -
    i
    \mathcal{A}^{\mu}_{aa}(\bm{k})h_{\bm{k}ab}^{\alpha_{1}}
    h_{\bm{k}ba}^{\alpha_{2}}
    \frac{f_{ab}}{\omega_{2}-\varepsilon_{\bm{k}ba}+i\gamma}
    +
    ih_{\bm{k}aa}^{\alpha_{1}}\mathcal{A}^{\mu}_{ab}(\bm{k})
    h_{\bm{k}ba}^{\alpha_{2}}
    \frac{f_{ab}}{\omega_{2}-\varepsilon_{\bm{k}ba}+i\gamma}\nonumber\\
    &\;\;\;\;\;\;
    \left.
    -
    i
    \mathcal{A}^{\mu}_{ac}(\bm{k})h_{\bm{k}cb}^{\alpha_{1}}
    h_{\bm{k}ba}^{\alpha_{2}}
    \frac{f_{ab}}{\omega_{2}-\varepsilon_{\bm{k}ba}+i\gamma}
    +
    ih_{\bm{k}ac}^{\alpha_{1}}\mathcal{A}^{\mu}_{cb}(\bm{k})
    h_{\bm{k}ba}^{\alpha_{2}}
    \frac{f_{ab}}{\omega_{2}-\varepsilon_{\bm{k}ba}+i\gamma}+
    [(\omega_{1},\alpha_{1})\longleftrightarrow(\omega_{2},\alpha_{2})]
    \right]\nonumber\\
    =&\,
    \frac{1}{2}
    \sum_{abc(a\neq c)}
    \int
    \frac{d^{d}\bm{k}}{(2\pi)^{d}}
    \int d\omega_{1}d\omega_{2}
    \left(
    \frac{-e^{3}}{\omega_{1}\omega_{2}}
    \right)
    E^{\alpha_{1}}(\omega_{1})E^{\alpha_{2}}(\omega_{2})
    \delta(\omega-\omega_{1}-\omega_{2})
    \nonumber\\
    &\times\left[
    i\mathcal{A}^\mu_{ca}(\bm{k})
    h_{\bm{k}ab}^{\alpha_{1}}
    h_{\bm{k}bc}^{\alpha_{2}}
    \frac
    {f_{ab}}
    {\omega_{1}-\varepsilon_{\bm{k}ab}+i\gamma}
    +
    \partial_{k^{\mu}}h_{\bm{k}ab}^{\alpha_{1}}
    h_{\bm{k}ba}^{\alpha_{2}}
    \frac{f_{ab}}{\omega_{2}-\varepsilon_{\bm{k}ba}+i\gamma}
    -
    i
    \mathcal{A}^{\mu}_{aa}(\bm{k})h_{\bm{k}ab}^{\alpha_{1}}
    h_{\bm{k}ba}^{\alpha_{2}}
    \frac{f_{ab}}{\omega_{2}-\varepsilon_{\bm{k}ba}+i\gamma}\right.\nonumber\\
    &\;\;\;\;\;\;
    \left.
    +
    i\mathcal{A}^{\mu}_{ab}(\bm{k})
    h_{\bm{k}aa}^{\alpha_{1}}
    h_{\bm{k}ba}^{\alpha_{2}}
    \frac{f_{ab}}{\omega_{2}-\varepsilon_{\bm{k}ba}+i\gamma}
    +
    i\mathcal{A}^{\mu}_{cb}(\bm{k})
    h_{\bm{k}ac}^{\alpha_{1}}
    h_{\bm{k}ba}^{\alpha_{2}}
    \frac{f_{ab}}{\omega_{2}-\varepsilon_{\bm{k}ba}+i\gamma}
    +
    [(\omega_{1},\alpha_{1})\longleftrightarrow(\omega_{2},\alpha_{2})]
    \right],
\end{align}
where we switched $a\leftrightarrow c$ in the first term. Similarly, the summation of $c$ is decomposed into the $c=b$ part and $c\neq b$ part. Then, for the $c\neq b$ part, the role of dummy variables $a$ and $b$ can be switched safely. Thus,
\begin{align}
    \langle\widehat{J}&^{\mu}_{\mathrm{shift}}\rangle^{(2)}(\omega)\nonumber\\
    =&\;
    \frac{1}{2}
    \sum_{abc(a\neq c,\,b\neq c)}
    \int
    \frac{d^{d}\bm{k}}{(2\pi)^{d}}
    \int d\omega_{1}d\omega_{2}
    \left(
    \frac{-e^{3}}{\omega_{1}\omega_{2}}
    \right)
    E^{\alpha_{1}}(\omega_{1})E^{\alpha_{2}}(\omega_{2})
    \delta(\omega-\omega_{1}-\omega_{2})
    \nonumber\\
    &\times\left[
    i\mathcal{A}^\mu_{ba}(\bm{k})
    h_{\bm{k}ab}^{\alpha_{1}}
    h_{\bm{k}bb}^{\alpha_{2}}
    \frac
    {f_{ab}}
    {\omega_{1}-\varepsilon_{\bm{k}ab}+i\gamma}
    +
    i\mathcal{A}^\mu_{ca}(\bm{k})
    h_{\bm{k}ab}^{\alpha_{1}}
    h_{\bm{k}bc}^{\alpha_{2}}
    \frac
    {f_{ab}}
    {\omega_{1}-\varepsilon_{\bm{k}ab}+i\gamma}
    +
    \partial_{k^{\mu}}h_{\bm{k}ab}^{\alpha_{1}}
    h_{\bm{k}ba}^{\alpha_{2}}
    \frac{f_{ab}}{\omega_{2}-\varepsilon_{\bm{k}ba}+i\gamma}\right.\nonumber\\
    &\;\;\;\;\;\;
    -
    i
    \mathcal{A}^{\mu}_{aa}(\bm{k})h_{\bm{k}ab}^{\alpha_{1}}
    h_{\bm{k}ba}^{\alpha_{2}}
    \frac{f_{ab}}{\omega_{2}-\varepsilon_{\bm{k}ba}+i\gamma}
    +
    i\mathcal{A}^{\mu}_{ab}(\bm{k})
    h_{\bm{k}aa}^{\alpha_{1}}
    h_{\bm{k}ba}^{\alpha_{2}}
    \frac{f_{ab}}{\omega_{2}-\varepsilon_{\bm{k}ba}+i\gamma}\nonumber\\
    &\;\;\;\;\;\;
    \left.
    +
    i\mathcal{A}^{\mu}_{bb}(\bm{k})
    h_{\bm{k}ab}^{\alpha_{1}}
    h_{\bm{k}ba}^{\alpha_{2}}
    \frac{f_{ab}}{\omega_{2}-\varepsilon_{\bm{k}ba}+i\gamma}
    +
    i\mathcal{A}^{\mu}_{cb}(\bm{k})
    h_{\bm{k}ac}^{\alpha_{1}}
    h_{\bm{k}ba}^{\alpha_{2}}
    \frac{f_{ab}}{\omega_{2}-\varepsilon_{\bm{k}ba}+i\gamma}
    +
    [(\omega_{1},\alpha_{1})\longleftrightarrow(\omega_{2},\alpha_{2})]
    \;
    \right]
    \nonumber\\
    =&\;
    \frac{1}{2}
    \sum_{abc(a\neq c,\,b\neq c)}
    \int
    \frac{d^{d}\bm{k}}{(2\pi)^{d}}
    \int d\omega_{1}d\omega_{2}
    \left(
    \frac{-e^{3}}{\omega_{1}\omega_{2}}
    \right)
    E^{\alpha_{1}}(\omega_{1})E^{\alpha_{2}}(\omega_{2})
    \delta(\omega-\omega_{1}-\omega_{2})
    \nonumber\\
    &\times\left[
    i\mathcal{A}^\mu_{ba}(\bm{k})
    h_{\bm{k}ab}^{\alpha_{1}}
    h_{\bm{k}bb}^{\alpha_{2}}
    \frac
    {f_{ab}}
    {\omega_{1}-\varepsilon_{\bm{k}ab}+i\gamma}
    +
    i\mathcal{A}^\mu_{ca}(\bm{k})
    h_{\bm{k}ab}^{\alpha_{1}}
    h_{\bm{k}bc}^{\alpha_{2}}
    \frac
    {f_{ab}}
    {\omega_{1}-\varepsilon_{\bm{k}ab}+i\gamma}
    +
    \partial_{k^{\mu}}h_{\bm{k}ab}^{\alpha_{1}}
    h_{\bm{k}ba}^{\alpha_{2}}
    \frac{f_{ab}}{\omega_{2}-\varepsilon_{\bm{k}ba}+i\gamma}\right.\nonumber\\
    &\;\;\;\;\;\;
    -
    i
    \mathcal{A}^{\mu}_{aa}(\bm{k})h_{\bm{k}ab}^{\alpha_{1}}
    h_{\bm{k}ba}^{\alpha_{2}}
    \frac{f_{ab}}{\omega_{2}-\varepsilon_{\bm{k}ba}+i\gamma}
    +
    i\mathcal{A}^{\mu}_{ab}(\bm{k})
    h_{\bm{k}aa}^{\alpha_{1}}
    h_{\bm{k}ba}^{\alpha_{2}}
    \frac{f_{ab}}{\omega_{2}-\varepsilon_{\bm{k}ba}+i\gamma}\nonumber\\
    &\;\;\;\;\;\;
    \left.
    +
    i\mathcal{A}^{\mu}_{bb}(\bm{k})
    h_{\bm{k}ab}^{\alpha_{1}}
    h_{\bm{k}ba}^{\alpha_{2}}
    \frac{f_{ab}}{\omega_{2}-\varepsilon_{\bm{k}ba}+i\gamma}
    -
    i\mathcal{A}^{\mu}_{ca}(\bm{k})
    h_{\bm{k}ab}^{\alpha_{1}}
    h_{\bm{k}bc}^{\alpha_{2}}
    \frac{f_{ab}}{\omega_{1}-\varepsilon_{\bm{k}ab}+i\gamma}
    +
    [(\omega_{1},\alpha_{1})\longleftrightarrow(\omega_{2},\alpha_{2})]
    \right]
    \nonumber\\
    =&\;
    \frac{1}{2}
    \sum_{ab}
    \int
    \frac{d^{d}\bm{k}}{(2\pi)^{d}}
    \int d\omega_{1}d\omega_{2}
    \left(
    \frac{-e^{3}}{\omega_{1}\omega_{2}}
    \right)
    E^{\alpha_{1}}(\omega_{1})E^{\alpha_{2}}(\omega_{2})
    \delta(\omega-\omega_{1}-\omega_{2})
    \nonumber\\
    &\times\left[
    \partial_{k^{\mu}}h_{\bm{k}ab}^{\alpha_{1}}
    h_{\bm{k}ba}^{\alpha_{2}}
    \frac{f_{ab}}{\omega_{2}-\varepsilon_{\bm{k}ba}+i\gamma}
    +
    i\mathcal{A}^\mu_{ba}(\bm{k})
    h_{\bm{k}ab}^{\alpha_{1}}
    h_{\bm{k}bb}^{\alpha_{2}}
    \frac
    {f_{ab}}
    {\omega_{1}-\varepsilon_{\bm{k}ab}+i\gamma}
    -
    i
    \mathcal{A}^{\mu}_{aa}(\bm{k})h_{\bm{k}ab}^{\alpha_{1}}
    h_{\bm{k}ba}^{\alpha_{2}}
    \frac{f_{ab}}{\omega_{2}-\varepsilon_{\bm{k}ba}+i\gamma}
    \right.\nonumber\\
    &\;\;\;\;\;\;
    \left.
    +
    i\mathcal{A}^{\mu}_{ab}(\bm{k})
    h_{\bm{k}aa}^{\alpha_{1}}
    h_{\bm{k}ba}^{\alpha_{2}}
    \frac{f_{ab}}{\omega_{2}-\varepsilon_{\bm{k}ba}+i\gamma}
    +
    i\mathcal{A}^{\mu}_{bb}(\bm{k})
    h_{\bm{k}ab}^{\alpha_{1}}
    h_{\bm{k}ba}^{\alpha_{2}}
    \frac{f_{ab}}{\omega_{2}-\varepsilon_{\bm{k}ba}+i\gamma}
    +
    [(\omega_{1},\alpha_{1})\longleftrightarrow(\omega_{2},\alpha_{2})]
    \right].
\end{align}
We write the $[(\omega_{1},\alpha_{1})\longleftrightarrow(\omega_{2},\alpha_{2})]$ part explicitly:
\begin{align}
    \langle\widehat{J}&^{\mu}_{\mathrm{shift}}\rangle^{(2)}(\omega)\nonumber\\
    =&\;
    \frac{1}{2}
    \sum_{ab}
    \int
    \frac{d^{d}\bm{k}}{(2\pi)^{d}}
    \int d\omega_{1}d\omega_{2}
    \left(
    \frac{-e^{3}}{\omega_{1}\omega_{2}}
    \right)
    E^{\alpha_{1}}(\omega_{1})E^{\alpha_{2}}(\omega_{2})
    \delta(\omega-\omega_{1}-\omega_{2})
    \nonumber\\
    &\times\left[
    \partial_{k^{\mu}}h_{\bm{k}ab}^{\alpha_{1}}
    h_{\bm{k}ba}^{\alpha_{2}}
    \frac{f_{ab}}{\omega_{2}-\varepsilon_{\bm{k}ba}+i\gamma}
    +
    i\mathcal{A}^\mu_{ba}(\bm{k})
    h_{\bm{k}ab}^{\alpha_{1}}
    h_{\bm{k}bb}^{\alpha_{2}}
    \frac
    {f_{ab}}
    {\omega_{1}-\varepsilon_{\bm{k}ab}+i\gamma}
    -
    i
    \mathcal{A}^{\mu}_{aa}(\bm{k})h_{\bm{k}ab}^{\alpha_{1}}
    h_{\bm{k}ba}^{\alpha_{2}}
    \frac{f_{ab}}{\omega_{2}-\varepsilon_{\bm{k}ba}+i\gamma}
    \right.\nonumber\\
    &\;\;\;\;\;\;
    +
    i\mathcal{A}^{\mu}_{ab}(\bm{k})
    h_{\bm{k}aa}^{\alpha_{1}}
    h_{\bm{k}ba}^{\alpha_{2}}
    \frac{f_{ab}}{\omega_{2}-\varepsilon_{\bm{k}ba}+i\gamma}
    +
    i\mathcal{A}^{\mu}_{bb}(\bm{k})
    h_{\bm{k}ab}^{\alpha_{1}}
    h_{\bm{k}ba}^{\alpha_{2}}
    \frac{f_{ab}}{\omega_{2}-\varepsilon_{\bm{k}ba}+i\gamma}
    \nonumber\\
    &\;\;\;\;\;\;
    +
    \partial_{k^{\mu}}h_{\bm{k}ab}^{\alpha_{2}}
    h_{\bm{k}ba}^{\alpha_{1}}
    \frac{f_{ab}}{\omega_{1}-\varepsilon_{\bm{k}ba}+i\gamma}
    +
    i\mathcal{A}^\mu_{ba}(\bm{k})
    h_{\bm{k}ab}^{\alpha_{2}}
    h_{\bm{k}bb}^{\alpha_{1}}
    \frac
    {f_{ab}}
    {\omega_{2}-\varepsilon_{\bm{k}ab}+i\gamma}
    -
    i
    \mathcal{A}^{\mu}_{aa}(\bm{k})h_{\bm{k}ab}^{\alpha_{2}}
    h_{\bm{k}ba}^{\alpha_{1}}
    \frac{f_{ab}}{\omega_{1}-\varepsilon_{\bm{k}ba}+i\gamma}
    \nonumber\\
    &\;\;\;\;\;\;
    \left.
    +
    i\mathcal{A}^{\mu}_{ab}(\bm{k})
    h_{\bm{k}aa}^{\alpha_{2}}
    h_{\bm{k}ba}^{\alpha_{1}}
    \frac{f_{ab}}{\omega_{1}-\varepsilon_{\bm{k}ba}+i\gamma}
    +
    i\mathcal{A}^{\mu}_{bb}(\bm{k})
    h_{\bm{k}ab}^{\alpha_{2}}
    h_{\bm{k}ba}^{\alpha_{1}}
    \frac{f_{ab}}{\omega_{1}-\varepsilon_{\bm{k}ba}+i\gamma}
    \right]\nonumber\\
    =&\;
    \frac{1}{2}
    \sum_{ab}
    \int
    \frac{d^{d}\bm{k}}{(2\pi)^{d}}
    \int d\omega_{1}d\omega_{2}
    \left(
    \frac{-e^{3}}{\omega_{1}\omega_{2}}
    \right)
    E^{\alpha_{1}}(\omega_{1})E^{\alpha_{2}}(\omega_{2})
    \delta(\omega-\omega_{1}-\omega_{2})
    \nonumber\\
    &\times\left[
    \partial_{k^{\mu}}h_{\bm{k}ab}^{\alpha_{1}}
    h_{\bm{k}ba}^{\alpha_{2}}
    \frac{f_{ab}}{\omega_{2}-\varepsilon_{\bm{k}ba}+i\gamma}
    +
    i\mathcal{A}^\mu_{ba}(\bm{k})
    h_{\bm{k}ab}^{\alpha_{1}}
    h_{\bm{k}bb}^{\alpha_{2}}
    \frac
    {f_{ab}}
    {\omega_{1}-\varepsilon_{\bm{k}ab}+i\gamma}
    -
    i
    \mathcal{A}^{\mu}_{aa}(\bm{k})h_{\bm{k}ab}^{\alpha_{1}}
    h_{\bm{k}ba}^{\alpha_{2}}
    \frac{f_{ab}}{\omega_{2}-\varepsilon_{\bm{k}ba}+i\gamma}
    \right.\nonumber\\
    &\;\;\;\;\;\;
    +
    i\mathcal{A}^{\mu}_{ab}(\bm{k})
    h_{\bm{k}aa}^{\alpha_{1}}
    h_{\bm{k}ba}^{\alpha_{2}}
    \frac{f_{ab}}{\omega_{2}-\varepsilon_{\bm{k}ba}+i\gamma}
    +
    i\mathcal{A}^{\mu}_{bb}(\bm{k})
    h_{\bm{k}ab}^{\alpha_{1}}
    h_{\bm{k}ba}^{\alpha_{2}}
    \frac{f_{ab}}{\omega_{2}-\varepsilon_{\bm{k}ba}+i\gamma}
    \nonumber\\
    &\;\;\;\;\;\;
    +
    h_{\bm{k}ba}^{\alpha_{1}}
    \partial_{k^{\mu}}
    h_{\bm{k}ab}^{\alpha_{2}}
    \frac{f_{ab}}{\omega_{1}-\varepsilon_{\bm{k}ba}+i\gamma}
    -
    i\mathcal{A}^\mu_{ab}(\bm{k})
    h_{\bm{k}aa}^{\alpha_{1}}
    h_{\bm{k}ba}^{\alpha_{2}}
    \frac
    {f_{ab}}
    {\omega_{2}-\varepsilon_{\bm{k}ba}+i\gamma}
    -
    i
    \mathcal{A}^{\mu}_{aa}(\bm{k})
    h_{\bm{k}ba}^{\alpha_{1}}
    h_{\bm{k}ab}^{\alpha_{2}}
    \frac{f_{ab}}{\omega_{1}-\varepsilon_{\bm{k}ba}+i\gamma}
    \nonumber\\
    &\;\;\;\;\;\;
    \left.
    -
    i\mathcal{A}^{\mu}_{ba}(\bm{k})
    h_{\bm{k}ab}^{\alpha_{1}}
    h_{\bm{k}bb}^{\alpha_{2}}
    \frac{f_{ab}}{\omega_{1}-\varepsilon_{\bm{k}ab}+i\gamma}
    +
    i\mathcal{A}^{\mu}_{bb}(\bm{k})
    h_{\bm{k}ba}^{\alpha_{1}}
    h_{\bm{k}ab}^{\alpha_{2}}
    \frac{f_{ab}}{\omega_{1}-\varepsilon_{\bm{k}ba}+i\gamma}
    \right]\nonumber\\
    =&\;
    \frac{1}{2}
    \sum_{ab}
    \int
    \frac{d^{d}\bm{k}}{(2\pi)^{d}}
    \int d\omega_{1}d\omega_{2}
    \left(
    \frac{-e^{3}}{\omega_{1}\omega_{2}}
    \right)
    E^{\alpha_{1}}(\omega_{1})E^{\alpha_{2}}(\omega_{2})
    \delta(\omega-\omega_{1}-\omega_{2})
    \nonumber\\
    &\times\left[
    \left(
    \partial_{k^{\mu}}h_{\bm{k}ab}^{\alpha_{1}}
    -
    i
    \mathcal{A}^{\mu}_{aa}(\bm{k})h_{\bm{k}ab}^{\alpha_{1}}
    +
    i\mathcal{A}^{\mu}_{bb}(\bm{k})
    h_{\bm{k}ab}^{\alpha_{1}}
    \right)
    h_{\bm{k}ba}^{\alpha_{2}}
    \frac{f_{ab}}{\omega_{2}-\varepsilon_{\bm{k}ba}+i\gamma}
    +[(\omega_{1},\alpha_{1})\longleftrightarrow(\omega_{2},\alpha_{2})]
    \right]
\end{align}
There remains only the $a\neq b$ part since it is proportional to $f_{ab}$. Thus \eqref{eq:h1berry} is simplified to $h_{\bm{k} ab}^\mu=i\varepsilon_{\bm{k}ab}\mathcal{A}^{\mu}_{ab}(\bm{k})$. Using the following shorthands,
\begin{align}
r^{\mu}_{\bm{k}ab}\equiv i \varepsilon_{\bm{k}ab} \mathcal{A}_{ab}^{\mu}(\bm{k}),\;\;
r^{\mu\alpha}_{\bm{k}ab}\equiv \left[\partial_{k_\mu} r^\alpha_{\bm{k}ab} -i \left(\mathcal{A}^\mu_{aa}(\bm{k})-\mathcal{A}^\mu_{bb}(\bm{k})\right)r^\alpha_{\bm{k}ab}\right],
\end{align}
we obtain the final form of the shift current:
\begin{align}
    \langle\widehat{J}^{\mu}_{\mathrm{shift}}\rangle^{(2)}(\omega)
    =&\;
    \frac{1}{2}
    \sum_{ab}
    \int
    \frac{d^{d}\bm{k}}{(2\pi)^{d}}
    \int d\omega_{1}d\omega_{2}
    \left(
    \frac{-e^{3}}{\omega_{1}\omega_{2}}
    \right)
    E^{\alpha_{1}}(\omega_{1})E^{\alpha_{2}}(\omega_{2})
    \delta(\omega-\omega_{1}-\omega_{2})
    \nonumber\\
    &\times\left[
    r^{\mu\alpha_{1}}_{\bm{k}ab}r^{\alpha_{2}}_{\bm{k}ba}
    \frac{f_{ab}}{\omega_{2}-\varepsilon_{\bm{k}ba}+i\gamma}
    +[(\omega_{1},\alpha_{1})\longleftrightarrow(\omega_{2},\alpha_{2})]
    \right].
\end{align}

\newpage
\section{Derivation of optical response using real-time formalism}\label{sec:Keldysh}
In this appendix, we derive the optical response of the electronic system using real-time formalism. We consider the full Hamiltonian $H=H_0+V_E$ given by Eq.~\eqref{eq:Hamiltonian_Taylor2}; $H_0$ is defined by Eq.~\eqref{eq:H_0} and $V_E$ is defined by Eq.~\eqref{eq:appendixVE}. 
Assume that the state is initialized as in thermal equilibrium without perturbation, $\rho_0$, at time $t=-\infty$, the expectation value of a general operator $O(t)$ is given by
\begin{align}
\langle O(t)\rangle &= \frac{\mathrm{Tr} \left[  \rho_0 U^\dagger(t,-\infty)  O_I(t) U(t,-\infty) \right]}{\mathrm{Tr}[\rho_0]}\\
&=\frac{\mathrm{Tr} \left[  \rho_0 U(-\infty,\infty)U(\infty,t)  O_I(t) U(t,-\infty) \right]}{\mathrm{Tr}[\rho_0]}\\
&=\frac{\mathrm{Tr} \left[  \rho_0 \mathcal{T} e^{-i \int_{\mathcal{C}} V_I(t')dt'}O_I(t)  \right]}{\mathrm{Tr}[\rho_0]}
\end{align}
where path $\mathcal{C}$ is from $- \infty \rightarrow \infty \rightarrow -\infty $ and the time evolution operator in interaction picture $U$ is given by
\begin{equation}
    U(t_1,t_2)=\exp \left(-i \int_{t_1}^{t_2} V_I(t') dt'\right)
\end{equation}
with
$V_I(t)\equiv e^{iH_0 t}V_E(t)e^{-iH_0 t}$. In the following, we use the notation $ \langle A\rangle _0 \equiv \mathrm{Tr}[\rho_0 A]/\mathrm{Tr}[\rho_0]$ for any operator $A$.

In the following, we set 
\begin{equation}
    O=v^\mu\equiv-i[r^\mu, H] =-i[r^\mu, H_0+V_E]  \label{eq:appendixrealtimev}
\end{equation}
corresponding to the current operator in the presence of electromagnetic fields. Here, for simplicity, we set $e=1,\hbar=1$ and omit the parameter $k$ and its integration for simplicity.

\subsection{Linear order}
Here, we assume the field $\bm{E}(t)= \bm{E}e^{-i\omega t+\gamma t}$, where $\gamma$ is a small positive constant, and the factor $e^{\gamma t}$ is added for the convergence at $t\rightarrow -\infty$. We are interested in the evolution of the operator from $t=-\infty \rightarrow t \sim 0$ to find the steady-state solution.
Note that the velocity operator given by Eq.~\eqref{eq:appendixrealtimev} also depends on the electric field. Expanding the above equation, we have
\begin{align}
\langle v^\mu(t)\rangle^{(1)}
&=\langle v_I^\mu(t)\rangle_0-i\langle v^\mu_I(t)  \int_{-\infty}^t dt_1 V_I(t_1) \rangle_0 - i \langle \int^{-\infty}_t dt_1 V_I(t_1) v^\mu_I(t)\rangle_0 \nonumber\\
&=e^{-i\omega t}\frac{iE^\alpha}{\omega}    f_ah^{\mu\alpha}_{aa}-\left(ie^{-i\omega t}\frac{iE^\alpha}{\omega}  f_a\int_{-\infty}^{t} dt_1 e^{+i\omega(t-t_1)} h^\mu_{ab} h^\alpha_{ba} e^{i\varepsilon_a(t- t_1)} e^{-i\varepsilon_b(t-t_1)} e^{\gamma t_1} - ( \mu\leftrightarrow \alpha) \right)\nonumber\\
&= e^{-i\omega t}\frac{iE^\alpha}{\omega} 
\left( f_ah^{\mu\alpha}_{aa}+ h^\mu_{ab}h^\alpha_{ba} \left[ \frac{f_{ab}}{(\omega-\varepsilon_{ba})+i\gamma}  \right]  \right)\label{eq:real1}
\end{align}
where
\begin{equation}
f_a \equiv \langle \psi_a| \rho_0 |\psi_a\rangle/\mathrm{Tr}[\rho_0],
\end{equation}
where $|\psi_a\rangle$ represents an eigenstate $a$.
We insert the complete set between the operators to obtain the second equation of Eq.~\eqref{eq:real1}.

\subsection{Second order}
Assume field $\bm{E}(t)= \bm{E}_1e^{-i\omega_1 t+\gamma t}+\bm{E}_2e^{-i\omega_2 t+\gamma t}$.
Expand the term including two fields and insert complete sets between every operator, similarly as in the case of linear order, we obtain
\begin{align}
\langle v^\mu(t)\rangle^{(2)}=&\langle v_I^\mu(t)\rangle_0-i\langle v^\mu_I(t)  \int_{-\infty}^t dt_1 V_I(t_1) \rangle_0 - i \langle \int^{-\infty}_t dt_1 V_I(t_1) v^\mu_I(t)\rangle_0  \nonumber\\
&-\langle v^\mu(t) \int_{-\infty}^{t} dt_2 \int^{t_2}_{-\infty}dt_1V_I(t_2)V_I(t_1)\rangle_0-\langle \int_t^{-\infty} d t_2 V_I(t_2) v^\mu(t) \int^{t}_{-\infty}dt_1V_I(t_1)\rangle_0 \nonumber\\
&-\langle \int^{-\infty}_{t_2} dt_1 \int_{t}^{-\infty}dt_2V_I(t_1)  V_I(t_2) v^\mu(t)\rangle_0 \label{eq:realtimesecond}
\end{align}

 For the first, second, and third terms in the right-hand side of Eq.~\eqref{eq:realtimesecond}, we obtain
\begin{align}
    (\text{first, second and third term})=&\frac{E^{\alpha}_1 E^{\beta}_2}{-\omega_1\omega_2} e^{-i\omega_1 t} e^{-i\omega_2 t} \left( h^{\mu\alpha}_{ab}h^{\beta}_{ba} \left[ \frac{f_{ab}}{(\omega_2-\varepsilon_{ba})+i\gamma}  \right] \right. \nonumber \\
    &+ \left.h^{\mu}_{ab}h^{\alpha\beta}_{ba} \left[ \frac{f_{ab}}{(\omega_1+\omega_2-\varepsilon_{ba})+2i\gamma}  \right]+h^{\mu\alpha \beta}_{aa} f_a \right)+(\alpha,\omega_1 \leftrightarrow \beta,\omega_2).
\end{align}

For the fourth term in the right-hand side of Eq.~\eqref{eq:realtimesecond}, we get
\begin{align}
(\text{fourth term})=&- h^\mu_{ab} h^{\alpha}_{bc} h^{\beta}_{ca}\frac{E^{\alpha}_1 E^{\beta}_2}{-\omega_1\omega_2} e^{-i\omega_1 t} e^{-i\omega_2 t}  f_a \nonumber\\
& \times\int_{-\infty}^{t} dt_2  \int_{-\infty}^{t_2}dt_1  e^{-i \omega_1(t_2-t)} e^{-i\omega_2 (t_1-t)}e^{\gamma (t_1+t_2)} e^{-i\varepsilon_b(t-t_2)} e^{-i\varepsilon_c(t_2-t_1)} e^{-i\varepsilon_a(t_1-t)}\nonumber\\
&+(\alpha,1\leftrightarrow \beta, 2) \nonumber\\
=&- \frac{h^\mu_{ab} h^{\alpha}_{bc} h^{\beta}_{ca}E^{\alpha}_1 E^{\beta}_2 e^{-i\omega_1 t} e^{-i\omega_2 t}  f_a}{[i(-\omega_2+\varepsilon_c-\varepsilon_a)+\gamma][i(-\omega_1-\omega_2-\varepsilon_a+\varepsilon_b)+\gamma]}+(\alpha,\omega_1\leftrightarrow \beta, \omega_2).
\end{align}
Performing a similar calculation for other terms, we arrive at the following result:
\begin{align}
\langle v^\mu(t)\rangle^{(2)}
=&\frac{E^{\alpha}_1 E^{\beta}_2}{-\omega_1\omega_2} e^{-i\omega_1 t} e^{-i\omega_2 t} \left( h^{\mu\alpha}_{ab}h^{\beta}_{ba} \left[ \frac{f_{ab}}{(\omega_2-\varepsilon_{ba})+i\gamma}  \right] +h^{\mu}_{ab}h^{\alpha\beta}_{ba} \left[ \frac{f_{ab}}{(\omega_1+\omega_2-\varepsilon_{ba})+2i\gamma}  \right]+h^{\mu\alpha \beta}_{aa} f_a \right) \nonumber\\
&+ h^\mu_{ab} h^{\alpha}_{bc} h^{\beta}_{ca}\frac{E^{\alpha}_1 E^{\beta}_2}{-\omega_1\omega_2} e^{-i\omega_1 t} e^{-i\omega_2 t}
\times \left( \frac{f_a}{[(\omega_2-\varepsilon_{ca})+i\gamma][(\omega-\varepsilon_{ba})+2i\gamma]}  \right. \nonumber \\
&\quad \quad -\frac{f_c}{[(\omega_2-\varepsilon_{ca})+i\gamma][\omega_1-\varepsilon_{bc})+i\gamma]} 
 \quad+\left.\frac{f_b}{[(\omega_1-\varepsilon_{bc})+i\gamma][(\omega-\varepsilon_{ba})+2i \gamma]} \right) \nonumber \\
&+(\alpha,\omega_1\leftrightarrow \beta, \omega_2) \nonumber \\
=&\frac{E^{\alpha}_1 E^{\beta}_2}{-\omega_1\omega_2} e^{-i\omega_1 t} e^{-i\omega_2 t} \left( h^{\mu\alpha}_{ab}h^{\beta}_{ba} \left[ \frac{f_{ab}}{(\omega_2-\varepsilon_{ba})+i\gamma}  \right] +h^{\mu}_{ab}h^{\alpha\beta}_{ba} \left[ \frac{f_{ab}}{(\omega_1+\omega_2-\varepsilon_{ba})+2i\gamma}  \right]+h^{\mu\alpha \beta}_{aa} f_a \right. \nonumber\\
& \quad \left.+h^\mu_{ab} h^{\alpha}_{bc} h^{\beta}_{ca} \left[ \frac{f_{ac} (\omega_1-\varepsilon_{bc}+i\gamma)+f_{bc}(\omega_2-\varepsilon_{ca}+i\gamma)}{(\omega_1-\varepsilon_{bc}+i\gamma)(\omega_2-\varepsilon_{ca}+i\gamma)(\omega-\varepsilon_{ba}+2i\gamma)} \right] \right) \nonumber\\
&+(\alpha,\omega_1\leftrightarrow \beta, \omega_2)
\end{align}
The results coincide with those derived using coherent path integral formalism presented in Appendix~\ref{sec:opticalresponse}.

\newpage
\section{Rice-mele model} \label{sec:Ricemale}
The Rice--Mele model \cite{RiceMele1982} is the simplest model capturing important features of conductivity with two input fields. It consists of a periodic array of unit cells, each containing two types of ions (A and B), arranged periodically with a single translational degree of freedom, so that the system is effectively one-dimensional and described by a single crystal momentum $k$. 
The electronic states are described within a tight-binding framework. The Hamiltonian is
\begin{align}
    H = \sum_{n} \left[
t_1\, c_{n,A}^\dagger c_{n,B}
+ t_2\, c_{n,B}^\dagger c_{n+1,A}
+ \mathrm{h.c.}
\right]
+ \sum_{n} \left(
V_A\, c_{n,A}^\dagger c_{n,A}
+ V_B\, c_{n,B}^\dagger c_{n,B}
\right),
\end{align}
where $t_1$ and $t_2$ are the hopping amplitudes, $\Delta = (V_A - V_B)/2$ is half the onsite energy difference, and $c_{n,\alpha}^\dagger$ ($c_{n,\alpha}$) creates (annihilates) an electron in sublattice $\alpha = A, B$ of the $n$th unit cell. 
Here, we assume that the A and B ions in a unit cell are assigned to the same spatial position, or are arranged in a zigzag configuration.
In the following, we set $t_1=t_2=t$ for simplicity.  Then, the inversion symmetry is broken when $V_A\neq V_B$.

In the momentum space, the Hamiltonian is represented by $2\times2$ matrix form:
\begin{align}
    H(k)=\bm{d}(k)\cdot\boldsymbol{\sigma},
\end{align}
where $\bm{\sigma}=\{\sigma_x,\sigma_y,\sigma_z\}$ with $\sigma_{x,y,z}$ denoting Pauli matrices, 
$d_{x}=t(1+\cos{ka}),
d_{y}=-t\sin{ka}$, and $
d_{z}=(V_{A}-V_{B})/2$ where $a$ is the lattice constant. 
The energy level and the corresponding Bloch state are
\begin{align}
    \varepsilon_{\pm}
    &=\pm |\bm{d}|,~~~~
    |u_{\pm}\rangle
    =
    \begin{pmatrix}
    \frac{d_{z}\pm|\bm{d}|}{d_{x}+id_{y}}\\
    1
    \end{pmatrix}.
\end{align}

\begin{figure}[htbp]
    \centering
    \includegraphics[width=0.5\linewidth]{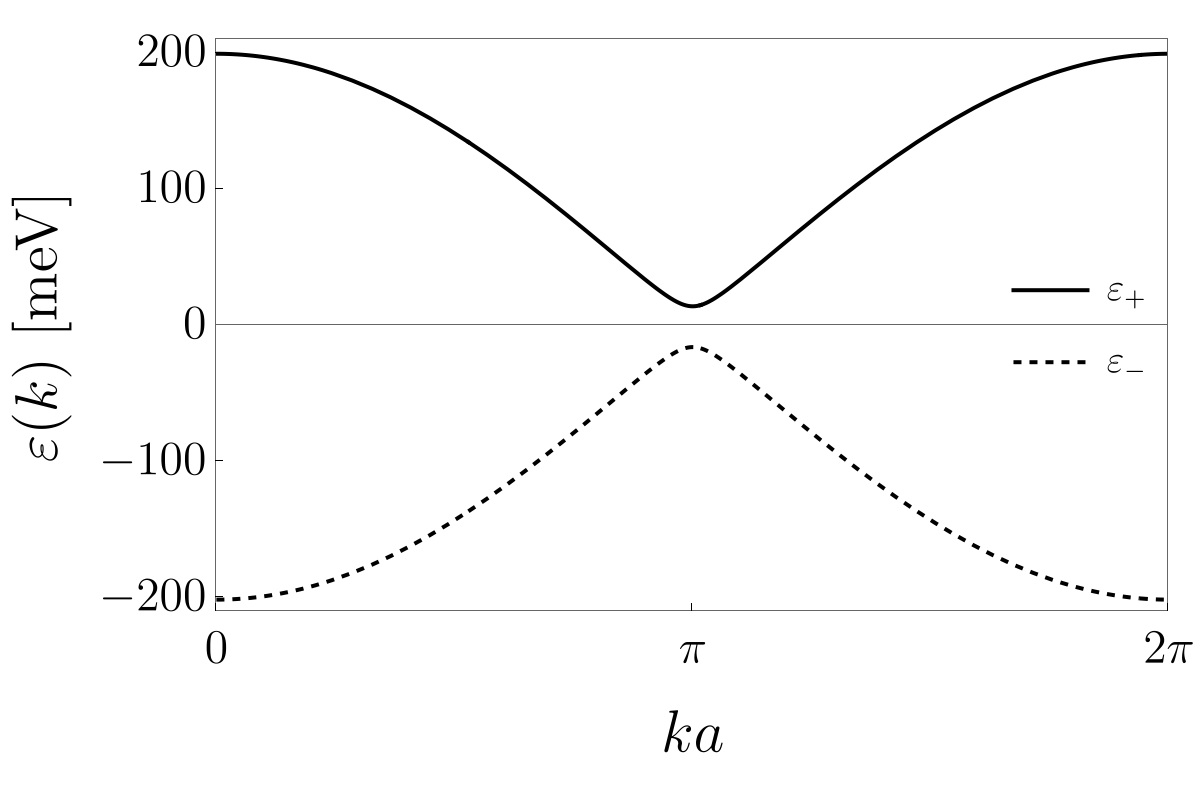}
    \caption{Band structure of Rice-Mele model. The potential of each site is $V_A=30 \ {\rm meV}, V_B=0 \ \rm meV$, while the hopping energy is $t=100 \ \rm meV$.  The spatial inversion symmetry is broken due to $V_A\neq V_B$, allowing the spatial shift of states.}
    \label{fig:Band}
\end{figure}

As a concrete example, we show the band structure in FIG.~\ref{fig:Band}. We have used the parameter set: $t=100 \ \rm meV$ and $V_A=30 \ {\rm meV}, V_B=0 \ \rm meV$, referring to the scale of typical electronic systems and topological insulators (Ref \cite{kim2017shift,osterhoudt2019colossal}). The band gap and the covering energy scale are determined by $|V_A-V_B|$ and $t$, respectively. 

\end{document}